\documentclass[preprint,12pt]{elsarticle}

\usepackage{amssymb}
\usepackage{graphicx}
\usepackage{epstopdf, epsfig}
\usepackage{lineno,hyperref}
\usepackage{booktabs} 
\usepackage{array} 
\usepackage{amsmath}
\usepackage{mathtools}
\usepackage{graphicx,caption,subcaption,color} 

\definecolor{jc}{rgb}{0.4,0.2,0.9}

\journal{Computers and Fluids}

\begin{document}

\begin{frontmatter}

\title{A comparative study of immersed boundary method and interpolated bounce-back scheme for no-slip boundary treatment in the lattice Boltzmann method: Part II, turbulent flows}

\author[label4,label1]{Cheng Peng\corref{cor1}}
\ead{cpengxpp@udel.edu}

\author[label2]{Orlando M. Ayala}
\ead{oayala@odu.edu}

\author[label5]{Jorge C\'esar Br\"andle de Motta}
\ead{jorge.brandle@coria.fr}

\author[label4,label1]{Lian-Ping Wang}
\ead{lwang@udel.edu}

\address[label4]{Department of Mechanics and Aerospace Engineering, Southern University of Science and Technology,
Shenzhen 518055, Guangdong, China}

\address[label1]{126 Spencer Lab, Department of Mechanical Engineering, University of Delaware, Newark, DE, USA, 19716}

\address[label2]{111A Kaufman Hall, Department of Engineering Technology, Old Dominion University, Norfolk, VA, 23529, USA}

\address[label5]{COmplexe de Recherche Interprofessionnel en A\'erothermochimie (CORIA), Universit\'{e} de Rouen Normandie, CNRS, INSA de Rouen, Saint-\'{E}tienne du Rouvray, France}

\cortext[cor1]{Corresponding author}

\begin{abstract}
In the first part of this study~\cite{peng2019}, we compared the performances of two categories of no-slip boundary treatments, {\it i.e.}, the interpolated bounce-back schemes and the immersed boundary methods in a series of laminar flow simulations within the lattice Boltzmann method. In this second part, these boundary treatments are further compared in the simulations of turbulent flows with complex geometry to provide a next-level assessment of these schemes. Two 
non-trivial turbulent flow problems, a fully developed turbulent pipe flow at a low Reynolds number, and a decaying homogeneous isotropic turbulent flow laden with a large number of resolved spherical particles are considered. The major problem of the immersed boundary method revealed by the present study is its incapability in computing the local velocity gradients inside the diffused interface, which can result in significantly underestimated dissipation rate and viscous diffusion locally near the particle surfaces. Otherwise, both categories of the no-slip boundary treatments are able to provide accurate results for most of turbulent statistics in both the carrier and dispersed phases, provided that sufficient grid resolutions are used.
\end{abstract}


\end{frontmatter}


\section{Introduction}
\label{sec:introduction}
In the first part of this study~\cite{peng2019}, we compared the performances of several representative interpolated bounce-back schemes and immersed boundary algorithms in several laminar flow configurations. In general, for the no-slip boundary treatment in the lattice Boltzmann method (LBM), the interpolated bounce-back (IBB) schemes can result in more accurate velocity, hydrodynamic force/torque, and dissipation rate calculations than the immersed boundary method (IBM). IBM, on the other hand, outperforms the IBB schemes in suppressing high-frequency numerical fluctuations in the instantaneous hydrodynamic force and torque results.

A very important application in developing accurate and efficient no-slip boundary treatments is the simulation and investigation of turbulent flows involving complex geometries. Particle-laden turbulent flows, for instance, are good examples encountered in many natural processes and engineering applications~\cite{balachandar2010turbulent}. In particular, when the size of the dispersed particles is not significantly smaller than the smallest flow length scale {\it i.e.}, the Kolmogorov length scale in a turbulent flow, the ability to resolve the no-slip boundary on the particle surfaces is a matter of ``life and death" in the investigation of these flows~\cite{prosperetti2015life}. While the laminar flow validations are useful as references to assess different no-slip boundary treatments to some extent, there are unique and genuine concerns in turbulent flows calling for additional validations and inter-comparisons 
going beyond the laminar flow tests.  Therefore, it is desirable to directly compare different boundary treatments in turbulent flow simulations. However, this type of rigorous inter-comparisons is largely missing.

There are two major difficulties in direct comparisons of no-slip boundary treatments in turbulent flow simulations. The first difficulty is the lack of reliable results for comparison. On the one hand, analytic results in turbulent flows are uncommon even without the presence of complex geometry. On the other hand, numerical benchmark results done with well established methods exist in the literature but the comparison becomes difficult when we focus on the details near the fluid-solid interfaces. The second major difficulty is the complexity of turbulent flows, which makes the isolation of the impact due to a specific factor difficult. Different implementations of the same method could also create divergences in results.

Due to the developments of both computational methods and large-scale supercomputer clusters, many interface-resolved particle-laden turbulent flow simulations have been reported since the last decade, and LBM has contributed to a significant part of these efforts~\cite{maxey2017simulation}. 
We note that both IBB schemes (adopted in~\cite{gao2013lattice,wang2016lattice,jebakumar2018lattice,peng2018direct}) and IBM (adopted in~\cite{ten2004fully,rubinstein2016lattice,eshghinejadfard2017immersed}) have been used for the no-slip boundary treatment on moving particle surfaces. While each of those studies has conducted certain validation tests and the results obtained in these particle-laden turbulent flow simulations can be trusted to some extent, direct comparisons among these boundary treatment methods for the same turbulent particle-laden flow are still largely unavailable. 
To the authors' knowledge, in turbulent flows, the only study involving direct comparisons among different numerical methods with resolved particles was reported recently by Br{\"a}ndle de Motta~{\it et al.}~\cite{de2019assessment}. In this work, three numerical methods, {\it i.e.}, the Lagrangian volume-of-fluid (VoF-Lag) method~\cite{vincent2014lagrangian}, the finite-volume based immersed boundary method (FV-IBM)~\cite{breugem2012second}, and LBM with interpolated bounce-back schemes (LBM-IBB)~\cite{wang2014study} are compared in a decaying homogeneous isotropic turbulence (HIT) laden with a few thousand of rigid spherical particles. Some obvious differences can be identified among the flow and particle statistics generated by each of these three methods, especially between the results of VoF-Lag and those from the other two methods. It is difficult to attribute these differences solely to the boundary treatments, since the flow solver used by these methods are different. In fact, even for the unladen single case, some differences were still identified.
The initial treatment of releasing particles to the flow field in each method could also contribute to the differences reported in Ref.~\cite{de2019assessment}, as differences were already observed from the very beginning of the simulation. Furthermore, since those simulations were conducted by different groups, it is difficult to ensure all the other simulation details, besides the no-slip boundary treatments, are handled identically. 
Therefore, how different no-slip boundary treatments impact the results of turbulent flow simulations remains an open question. 

In this paper, we will compare the performances of the two categories of boundary treatments in LBM, {\it i.e.}, IBM and IBB schemes, in two turbulent flow simulations. These boundary treatments are implemented in the same base code, reducing the sources of discrepancies. The first simulation is a direct numerical simulation of a single-phase turbulent pipe flow at a low Reynolds number. This case is chosen because published datasets based on other methods with grid meshes in a body-fitted cylindrical coordinate (e.g.,~\cite{Loulou97,El2013} based on spectral methods, and~\cite{Wagner01} based on finite-volume method) are available as benchmark results and the performances of the boundary treatments can be examined with the presence of a curved boundary. The second case is the particle-laden decaying HIT studied in Br{\"a}ndle de Motta {\it et al.}~\cite{de2019assessment}. Here we conduct comparisons with better control of the implementation details. To quantitatively examine the results, the second case will be performed using two different grid resolutions.

The paper is arranged as follows. In Sec. 2, a brief introduction of the first test problem and the simulation setup is given. Some important implementation details is also discussed, followed by
the simulation results and inter-comparisons of IBB and IBM.
In Sec. 3, the simulation results of the particle-laden decaying HIT  are presented and compared. At last, the key conclusions and additional remarks are presented in Sec. 4.

\section{Direct numerical simulations of a turbulent pipe flow}

\subsection{Problem description and simulation setup}

\begin{figure}
\centering
\includegraphics[width=120mm]{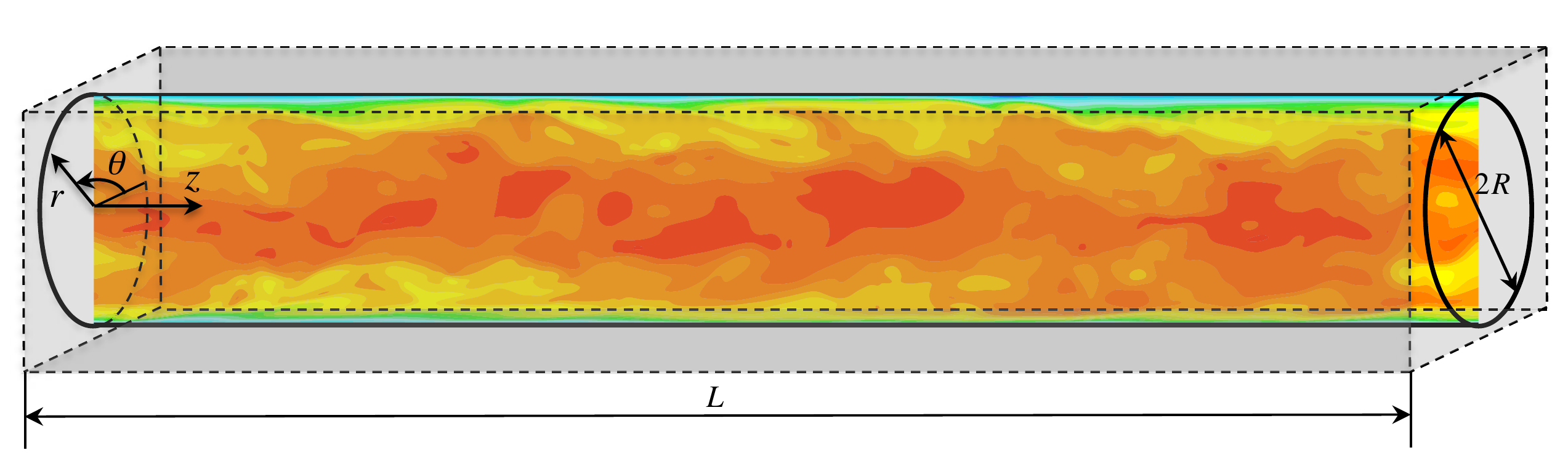}
\caption{A sketch of a turbulent pipe flow.}
\label{fig:turbulentpipeflow}
\end{figure}

The first case we consider is the direct numerical simulations of a turbulent pipe flow at a given Reynolds number. The sketch of this flow is shown in Fig.~\ref{fig:turbulentpipeflow}, where $z$, $r$, and $\theta$ represent the streamwise, radial, and azimuthal direction, respectively. The radius of the pipe is $R$, the length of the pipe is $L$. The periodic boundary condition is assumed in the streamwise direction. On the pipe wall, the no-slip boundary condition is applied. The flow is driven by a constant body force $\rho g$ per unit volume. At the stationary state, the balance between the driving force and the viscous drag provided by the pipe wall is established as $2\pi RL \langle \tau_{w} \rangle = \pi R^2 L\rho g$, where $\langle \tau_{w}\rangle$ is the wall shear stress $\tau_w = \left. \mu \frac{\partial u_z}{\partial r} \right|_{r=R}$, averaged over the time and the cylindrical solid-fluid interface. 
In a wall-bounded turbulent flow, the thin layer attached to the wall where the viscous effects are dominating is called the viscous sublayer. The characteristic velocity and length scale in the viscous sublayer are the friction velocity $u_{\tau} = \sqrt{ \langle \tau_{w} \rangle /\rho}$ and the wall unit $y_{\tau} = \nu/u_{\tau}$, respectively, where $\nu$ is the kinematic viscosity of the flow. $u_{\tau}$ and $y_{\tau}$ are also often referred as the inner scale of wall-bounded turbulence. A friction Reynolds number can be defined with the friction velocity as $Re_{\tau} = u_{\tau}R/\nu = R/y_{\tau}$. In this study we set $Re_{\tau} = 180$, which corresponds to a bulk Reynolds number $Re_{bulk} = 2UR/\nu\approx 5300$, where $U$ is the mean flow velocity averaged over the whole pipe at the statistically stationary stage. A turbulent pipe flow is expected when $Re_{bulk}\ge 2300\sim3000$.  

In order to compare the performances of IBB and IBM, one IBB scheme, {\it i.e.}, the linear IBB scheme proposed by Yu et al.~\cite{yu2003viscous} (labeled ``IBB") and two IBM algorithms, {\it i.e.}, the IBM proposed by Uhlmann~\cite{uhlmann2005immersed} (labeled ``IBM-U'') and the IBM improved by Breugem~\cite{breugem2012second} (labeled ``IBM-B'') with a retraction distance of $R_{d} = 0.4\delta x$ ($\delta x$ is the grid spacing) being applied to the Lagrangian grid points, are used to implement the no-slip condition on the pipe wall. 
Detailed descriptions of these schemes can be found in the cited references~\cite{yu2003viscous,uhlmann2005immersed,breugem2012second} and the first part of this study~\cite{peng2019}. In fact, although Breugem's IBM has been improved based on Uhlmann's original IBM in three aspects, the two IBM algorithms tested in the present case have only one difference, that is the former has a non-zero retraction distance $R_{d} = 0.4\delta x$ while the latter has a zero retraction distance. The delta-function used to exchange information between the Lagrangian and Eulerian grids is the four-point delta-function designed by Peskin~\cite{peskin2002immersed}. We chose this delta-function mainly because it ensures numerical stability in the various laminar flow tests discussed in the first part of this study, compared to the three-point delta-function and linear two-point delta-function~\cite{peng2019}. The retraction distance $R_{d} = 0.4\delta x$ is adopted because it provided the most accurate results in the laminar flow tests, when the four-point delta-function was employed. The cell volume of the Lagrangian grid in the IBM is chosen to be approximately equal to the cell volume of the Eulerian grid, as suggested by Uhlmann~\cite{uhlmann2005immersed}. In this case, the Lagrangian grid points are generated in a two-dimensional circle then copied in the streamwise direction. The streamwise locations of the Lagrangian grid points are identical with the streamwise locations of the Eulerian grid points. In order to ensure the no-slip boundary condition on the Lagrangian grids, we iterate the boundary force twice (excluding the first prediction) as suggested by Breugem~\cite{breugem2012second}. A larger number of iteration steps may further improve the no-slip boundary enforcement, but it is not computationally efficient.  

The pipe-flow simulation results with Yu et al.'s linear IBB scheme have been reported in our recent publication~\cite{peng2018directb}. We did not repeat the simulation but directly used the published results for comparison for the sake of saving computational resources. The linear interpolation scheme rather than its quadratic counterpart is chosen because the quadratic interpolation scheme did not yield a stable simulation in this turbulent pipe flow simulation. 

Both the IBB and the IBM simulations used the same flow solver based on a D3Q27 MRT LB model, but run with a single relaxation parameter to minimize parameter dependence. For more details of the simulations, such as how to set up initial condition, how to accelerate the transition from laminar flow to turbulent flow, and how to compute the turbulence statistics in cylindrical coordinates, readers can refer to Ref.~\cite{peng2018directb}. Here we only recapitulate the key parameters of the two simulations in Table~\ref{tab:turbulentpipe}. A slightly finer grid resolution $R = 157\delta x$ is used in the IBM simulation than $R = 148.5\delta x$ used in the IBB simulation. The slightly different grid resolutions were used mainly for the convenience to decompose the computational domain. In the IBB simulation, a layer of grid points outside the pipe wall is needed to temporarily store the distribution functions for interpolated bounce-back. A pipe radius of $R = 148.5\delta x$ can be fitted into a cross section of the computational domain of $N_x\times N_y = 300\times 300$. In the IBM simulations, on the other hand, since the four-point delta-function is used, two layers of grid points outside the pipe are required. Instead of reducing the pipe radius, the computational domain is expanded to $N_x\times N_y = 320\times 320$. To maximize the effective region for the flow domain, the pipe radius is set to $R = 157\delta x$ in the IBM simulations.
Since the grid resolutions in the IBB simulation and in the two IBM simulations are not too different, the results can be fairly compared.

\begin{table*}
\caption{Physical and numerical parameters used for the simulation of turbulent pipe flow. All the parameters are given in lattice units, {\it i.e.}, $\delta_{x} = \delta_{y}= \delta_{z}= 1$. From the left to right column: pipe geometry, grid resolution, viscosity, friction velocity, friction Reynolds number, relative grid resolution, retraction distance of Lagrangian grid points in the IBM simulation.}
\footnotesize
\begin{center}
\begin{tabular}
{c c c c c c c c}
\toprule
 &$R\times L$&$N_{x}\times N_{y}\times N_{z}$&$\nu$&$u_{\tau}$&$Re_{\tau}$&$\delta x/y_{\tau}$&$r_{d}$\\
\midrule 
 IBB&$148.5\times1799$&$300\times300\times1799$&0.0032&0.00388&180&1.212& \\
 IBM-U&$157\times1919$&$320\times320\times1919$&0.0032&0.00367&180&1.146&0.0\\
IBM-B&$157\times1919$&$320\times320\times1919$&0.0032&0.00367&180&1.146&0.4\\
\bottomrule
\end{tabular} 
\end{center}
\label{tab:turbulentpipe}
\end{table*} 

\subsection{Results and discussions}

\subsubsection{Mean flow statistics}

\begin{figure}
\centering
\includegraphics[width=90mm]{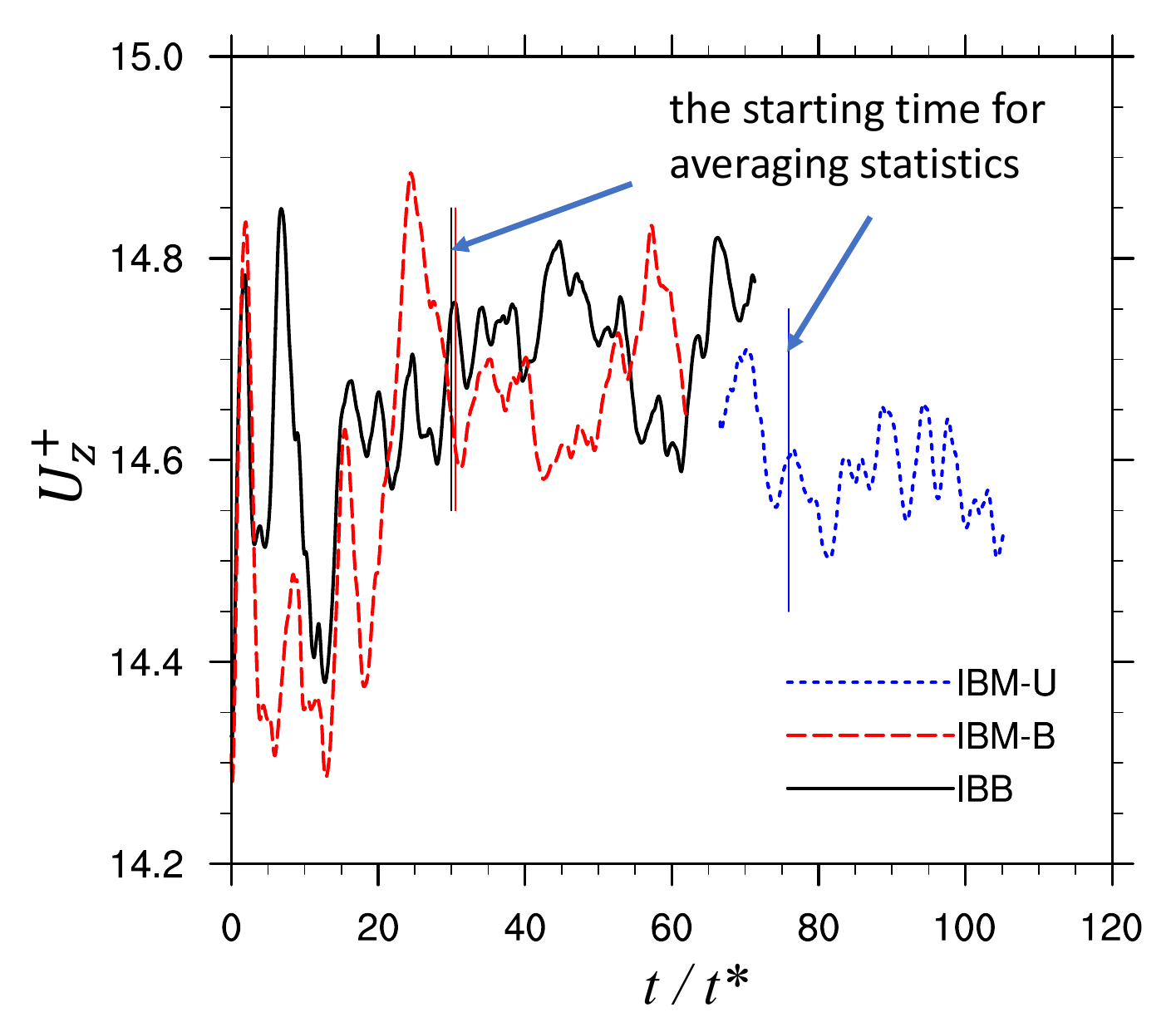}
\caption{Time evolution of the bulk flow velocity. The three vertical lines show the starting time for statistics gathering in each simulation. }
\label{fig:timeevolution}
\end{figure}

The evolution of the bulk flow velocity, {\it i.e.}, the flow velocity averaged over the whole pipe volume, in the three simulations are shown in Fig.~\ref{fig:timeevolution}. Both the IBB and the IBM-B simulations are started from the same initial laminar velocity field. We applied a perturbation force field (see Ref.~\cite{peng2018directb} for details) in the initial 3 eddy turnover times (the eddy turnover time $t^{*}$ is defined as $t^* = R/u_{\tau}$), so the transition from the laminar to turbulent flow could happen relatively fast. After about 25 eddy turnover times from the initial time, both the IBB and the IBM-B simulations reached a statistically stationary state. The IBM-U simulation was started from the velocity field of IBM-B simulation at $t/t^*=64.3$. This initialization was done in order to save computing resources. A new statistical stationary state in the IBM-U simulation is established after about 10 eddy turnover times, as shown in Fig.~\ref{fig:timeevolution}. The turbulence statistics examined in the present study are averaged over 30 eddy turnover times and for roughly 1300 time frames in each simulation. For the IBB, IBM-B, and IBM-U simulations, the bulk flow velocities at the stationary state are $14.72\pm0.057$, $14.67\pm0.060$, and $14.58\pm0.040$, respectively, when normalized by the corresponding friction velocity. The value after each $\pm$ is the standard derivation of the corresponding mean streamwise velocity.

\begin{figure}
\centering
\includegraphics[width=90mm]{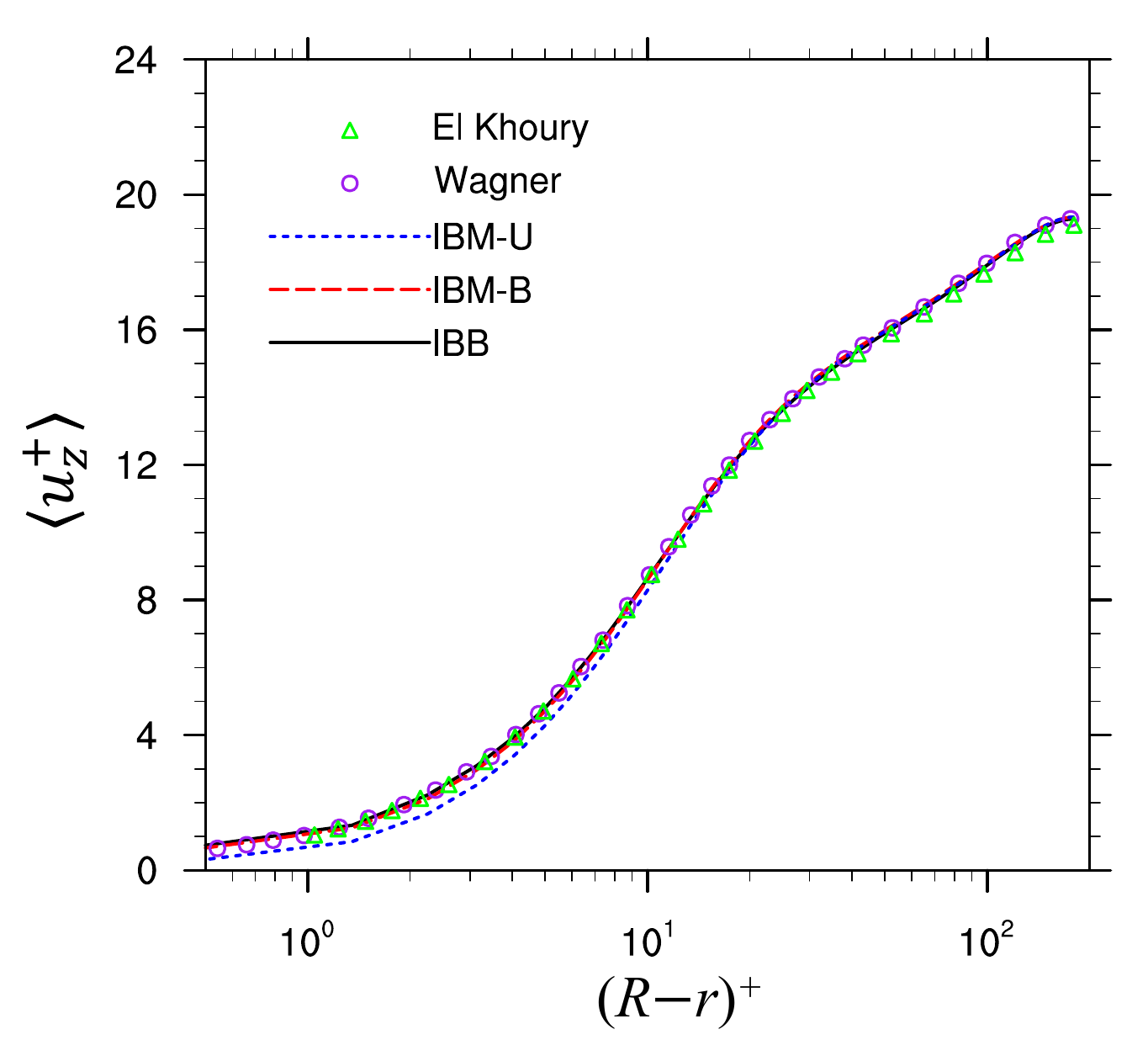}
\caption{Mean streamwise flow velocity at the stationary state, as a function of the distance from the pipe wall. Results from~\cite{Wagner01}~and~\cite{El2013} are given for comparison.}
\label{fig:meanflowvelocity}
\end{figure}

\begin{figure}
\centering
\includegraphics[width=90mm]{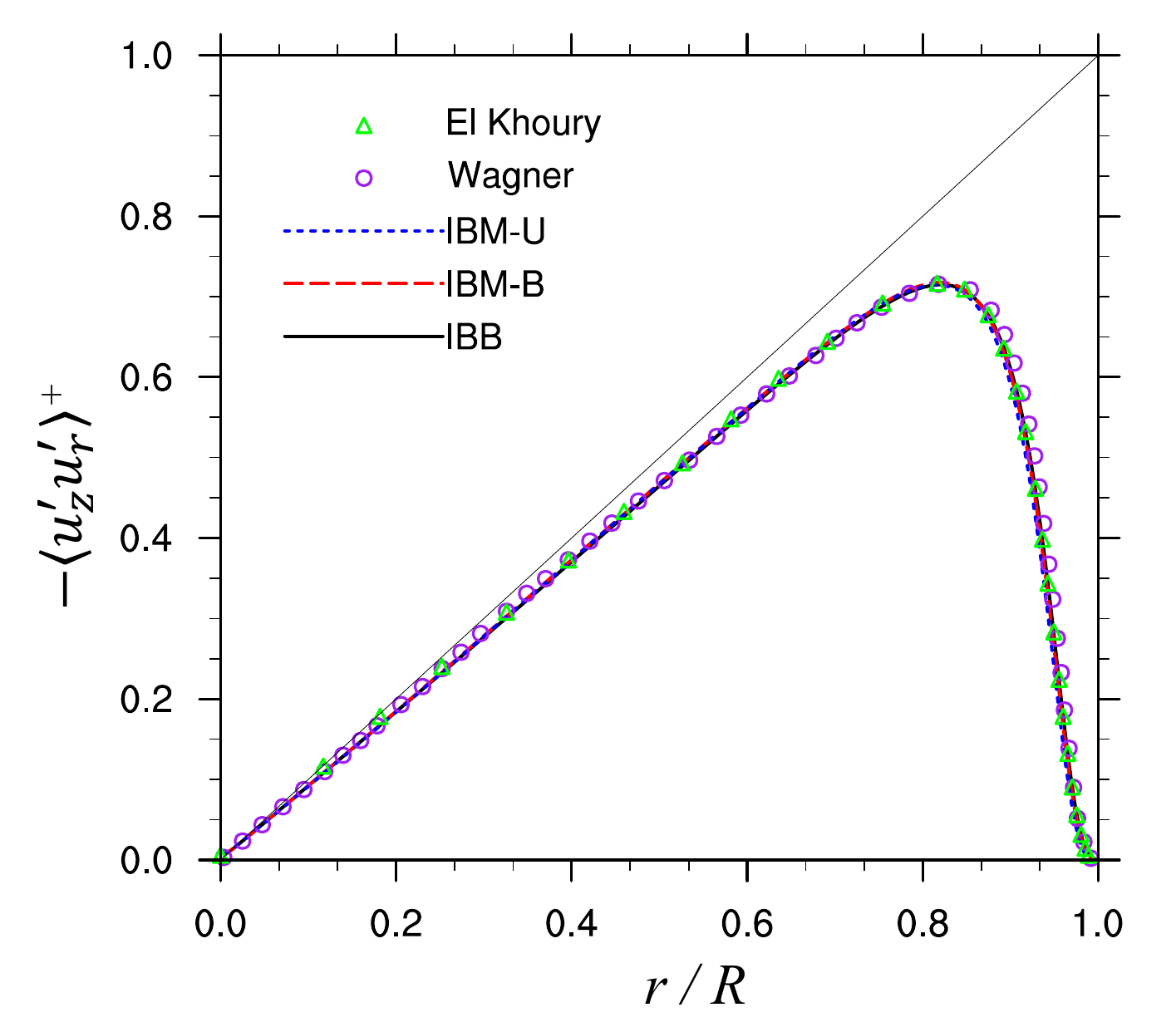}
\caption{The Reynolds stress as a function of the distance from the pipe wall. Results from~\cite{Wagner01}~and~\cite{El2013} are given for comparison.}
\label{fig:Reynoldsstress}
\end{figure}

The mean streamwise flow velocity profiles at the stationary state, $\langle u_z^+\rangle$, are presented in Fig.~\ref{fig:meanflowvelocity}, together with the benchmark results, extracted from the spectral simulation~\cite{El2013} and finite-volume simulation~\cite{Wagner01}, both conducted in cylindrical coordinates on non-uniform grids and with the identical friction Reynolds number.
 In this section, $\langle\cdots\rangle$ indicates the ensemble average over the two homogeneous directions, $z$ and $\theta$, and time. Details on how to transform data from the Cartesian coordinates on which the present LBM simulations are based, to cylindrical coordinates, can be also found in~\cite{peng2018directb}. The profiles of the IBB simulation and the IBM-B simulation match excellently with the benchmark results, but the profile of the IBM-U simulation is visibly below the benchmark results in the near wall region of $(R-r)^{+}\le 10$. This is because the skin drag force predicted by the immersed pipe wall to the flow is usually overrated, and retracting the Lagrangian grid tends to offset this overestimation~\cite{breugem2012second,peng2019}.  
While IBM only possesses a first-order accuracy in the no-slip boundary treatment, given a sufficient grid resolution and the retraction of Lagrangian grids, the simulated mean flow velocity is quite reliable.

\begin{figure}
\centering
\includegraphics[width=90mm]{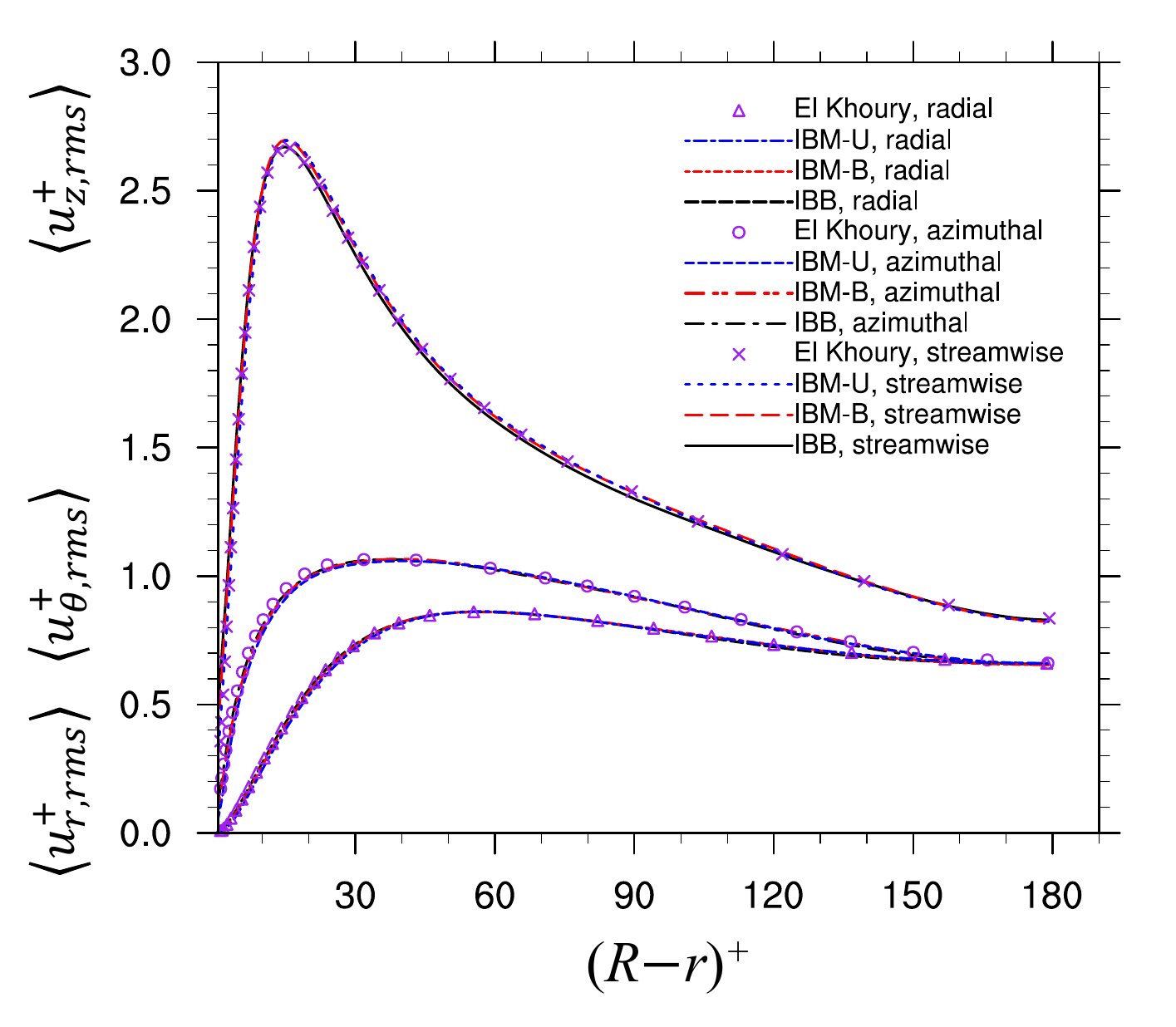}
\caption{r.m.s. velocity as a function of the distance from the pipe wall. Results from~\cite{El2013} are given for comparison.}
\label{fig:rmsvelocity}
\end{figure}

\subsubsection{Turbulent intensity statistics}

To go further in the inter-comparison, we now study the turbulent intensity. The profiles of the Reynolds stress and the root-mean-square (r.m.s.) velocities in all three spatial directions are shown in Fig.~\ref{fig:Reynoldsstress} and Fig.~\ref{fig:rmsvelocity}, respectively. The profiles from all three  simulations not only collapse, but also match very well with the benchmark results.
The retraction of Lagrangian grid points does not result in significant improvements on the simulated turbulent intensity. This indicates that the over-prediction of the drag force by IBM mainly affects the mean flow, and this does not impact the turbulent fluid motion. 

\begin{figure}
\centering
\includegraphics[width=90mm]{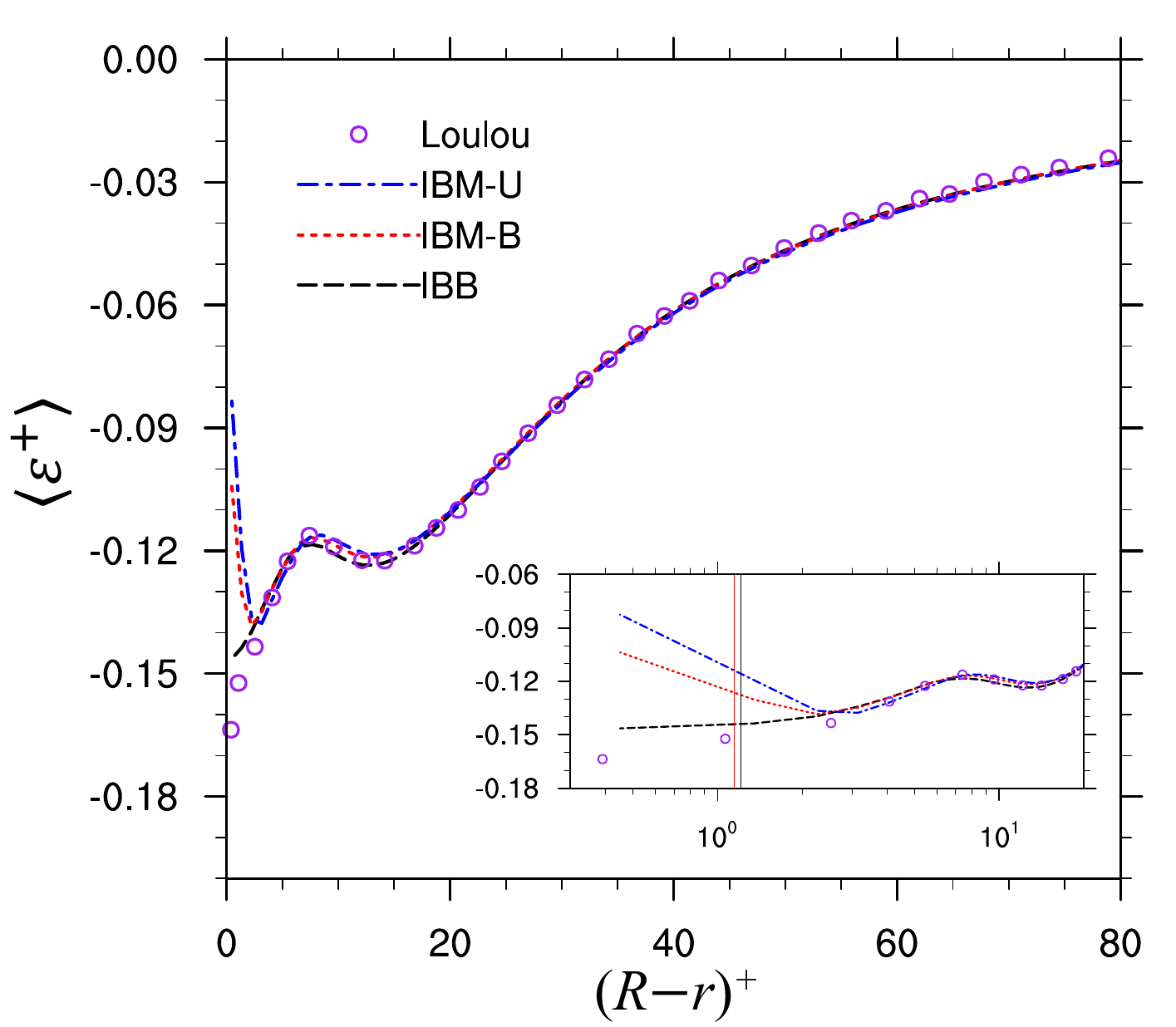}
\caption{Averaged dissipation rate as a function of the distance from the pipe wall. Results from~\cite{Loulou97} are given for comparison. The inserted plot is a zoom-in exhibition of the results near the wall. The red and black vertical lines indicates the radial location of $\delta x$ in the IBM and IBB simulations, respectively.}
\label{fig:dissipationrate}
\end{figure}

The profiles of the averaged viscous dissipation rate of the turbulent kinetic energy (TKE), defined as $\varepsilon = 2\mu s_{ij}'s_{ij}'$, where $\mu$ is the dynamic viscosity, $s_{ij}'= \left( \frac{1}{2}\left[\frac{\partial u_i}{\partial x_j} + \frac{\partial u_j}{\partial x_i}\right]\right)'$ is the fluctuation part of strain rate tensor, are shown in Fig.~\ref{fig:dissipationrate}.  The results are compared to the spectral simulation of Loulou {\it et al}.~\cite{Loulou97} at a slightly larger friction Reynolds number $Re_{\tau} = 190$. In the first part of this study, we had already pointed out that due to the presence of a non-zero boundary force term in IBM, the computation of the dissipation rate inside the diffused fluid-solid interface is no longer accurate. The boundary force term in IBM alters the governing equation inside the diffused interface, {\it i.e.}, the actual equation solved by IBM is no longer identical to the equation governing the physical flow, which affects the calculation of the local velocity gradients.  This observation also applies to the dissipation rate of TKE. As shown in Fig.~\ref{fig:dissipationrate}, close to the pipe wall, dissipation rates of the two IBM simulations not only are significantly underestimated, but also exhibit an unphysical decrease approaching the pipe wall. 

On the other hand, although the dissipation rate calculated from the IBB simulation also deviates from the benchmark result to some extent, its monotonically increasing slope towards the pipe wall is successfully captured. The deviation from the benchmark results is likely the outcome of insufficient grid resolution to fully resolve the small eddy structures locally. 
The size of the smallest eddy structure, {\it i.e.}, the Kolmogorov length scale $\eta$ is related to the local dissipation rate as $\eta^{+} = (\varepsilon^{+})^{-1/4}$. In the present simulations, the grid resolution was chosen so it is sufficient to resolve the Kolmogorov length based on the averaged dissipation rate near the wall. However, due to the intermittency in the turbulent flow, this grid resolution may become inadequate for certain instantaneous small-scale fluid motion. In the benchmark simulation conducted by Loulou et al.~\cite{Loulou97}, a non-uniform grid in the radial direction was used, with the finest grid resolution of $\delta r^{+} = 0.39$ at the wall. In the present LBM simulations, however, a uniform Cartesian grid was adopted, the radial grid resolution at the wall is $\delta r^{+} = 1.21$ for the IBB simulation and $\delta r^{+} = 1.15$ for the IBM simulations.

To conclude this study, we examine the balance in the budget equations of component-wise TKE. These equations in cylindrical coordinates can be derived as 

\begin{subequations}
\begin{align}
\begin{split}
    &0=\underbrace{\langle -u_{z}'u_{r}'\rangle\frac{\partial \langle u_{z} \rangle}{\partial r}}_{E_{Oz}}\underbrace{ - \frac{1}{r}\frac{\partial}{\partial r}\left(r\frac{1}{2}\langle u_{z}'u_{z}'u_{r}'\rangle\right)}_{E_{Tz}}\underbrace{+\frac{1}{\rho}\bigg\langle p'\frac{\partial u_{z}'}{\partial z}\bigg\rangle}_{E_{Pz}} \\
    &\underbrace{+ \frac{2\nu}{r}\frac{\partial}{\partial r}\left(r\langle u_{z}'s_{zr}'\rangle\right)}_{E_{Vz}}\underbrace{ - 2\nu\left(\bigg\langle s_{zr}'\frac{\partial u_{z}'}{\partial r}\bigg\rangle + \bigg\langle s_{z\theta}'\frac{1}{r}\frac{\partial u_{z}'}{\partial \theta}\bigg\rangle +\bigg\langle s_{zz}'\frac{\partial u_{z}'}{\partial z}\bigg\rangle  \right)}_{E_{Dz}},
    \end{split}
    \label{eq:streamwiseTKE}\\
    \begin{split}
        &0=\underbrace{-\frac{1}{r}\frac{\partial}{\partial r}\left(r\frac{1}{2}\langle u_{r}'u_{r}'u_{r}'\rangle\right)}_{E_{Tr}} \underbrace{-\frac{\partial}{\partial r}\langle u_{r}'p' \rangle + \frac{1}{\rho}\bigg\langle p'\frac{\partial u_{r}'}{\partial r}\bigg\rangle}_{E_{Pr}} \underbrace{+ \frac{2\nu}{r}\frac{\partial}{\partial r}\left(r\langle u_{r}'s_{rr}'\rangle\right)}_{E_{Vr}} \\
        &\underbrace{-2\nu\left(\langle s_{rr}'\frac{\partial u_{r}'}{\partial r}\rangle +\langle s_{r\theta}'\frac{1}{r}\frac{\partial u_{r}'}{\partial \theta}\rangle + \langle s_{rz}'\frac{\partial u_{r}'}{\partial z}\rangle \right)}_{E_{Dr}}
    \end{split}
    \label{eq:radialTKE}\\
    \begin{split}
    &0 = \underbrace{-\frac{1}{r}\frac{\partial}{\partial r}\left(r\frac{1}{2}\langle u_{\theta}'u_{\theta}'u_{r}'\rangle\right)}_{E_{T\theta}} \underbrace{+\bigg\langle p'\frac{1}{r}\frac{\partial u_{\theta}'}{\partial \theta}\bigg\rangle}_{E_{P\theta}} \underbrace{+ \frac{2\nu}{r}\frac{\partial}{\partial r}\left(r\langle u_{\theta}'s_{\theta r}'\rangle\right)}_{E_{V\theta}} \\
    &\underbrace{- 2\nu\left(\langle s_{\theta\theta}'\frac{1}{r}\frac{\partial u_{\theta}'}{\partial \theta}\rangle + \langle s_{\theta r}'\frac{\partial u_{\theta}'}{\partial r}\rangle + \langle s_{\theta z}'\frac{\partial u_{\theta}'}{\partial z}\rangle \right)}_{E_{D\theta}}
    \end{split}
    \label{eq:azimuthalTKE}
\end{align}
\label{eq:TKEbudget}
\end{subequations}
 
\begin{figure}
\centering
\includegraphics[width=110mm]{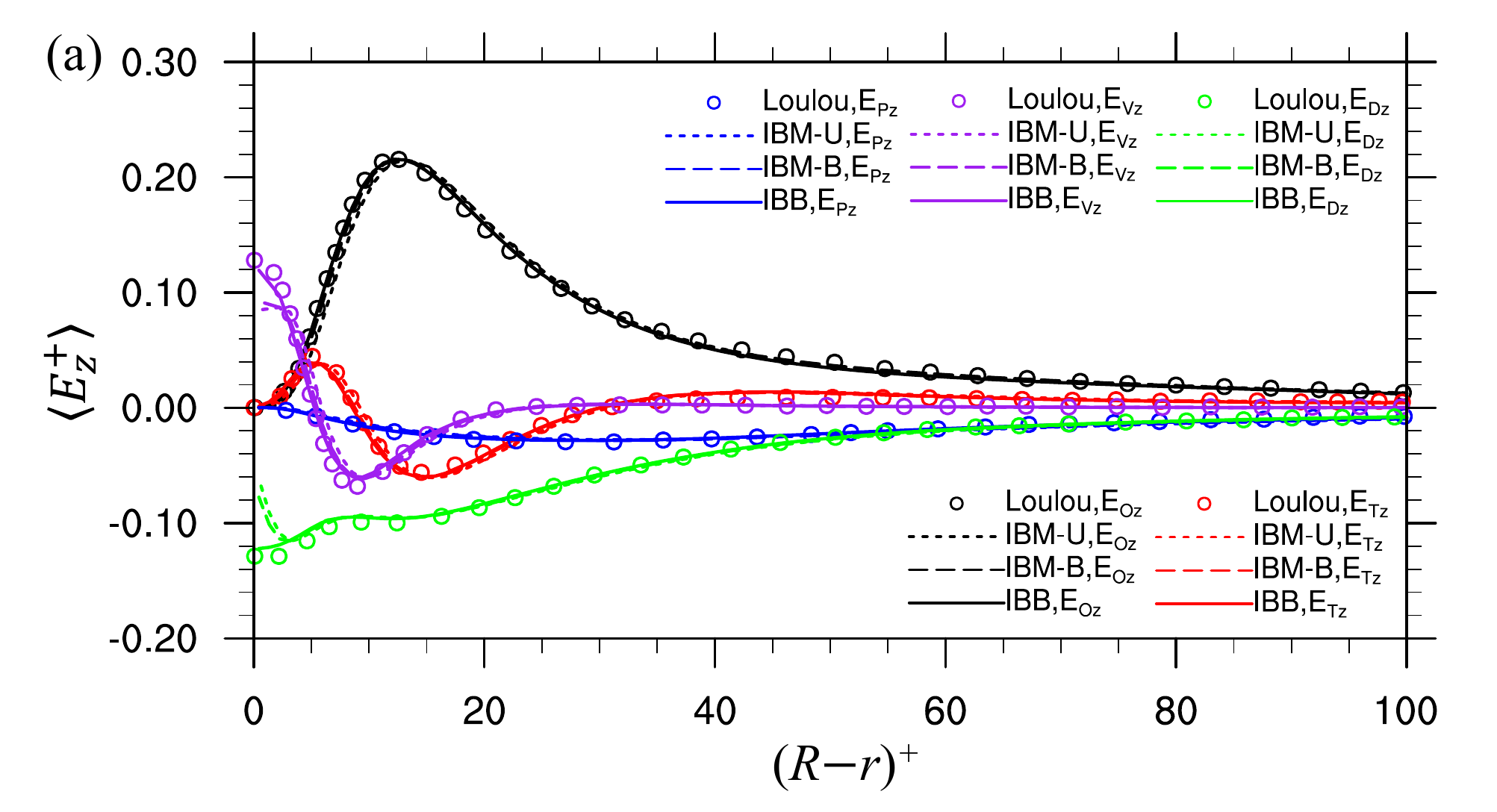}
\vspace{0.1in}
\includegraphics[width=110mm]{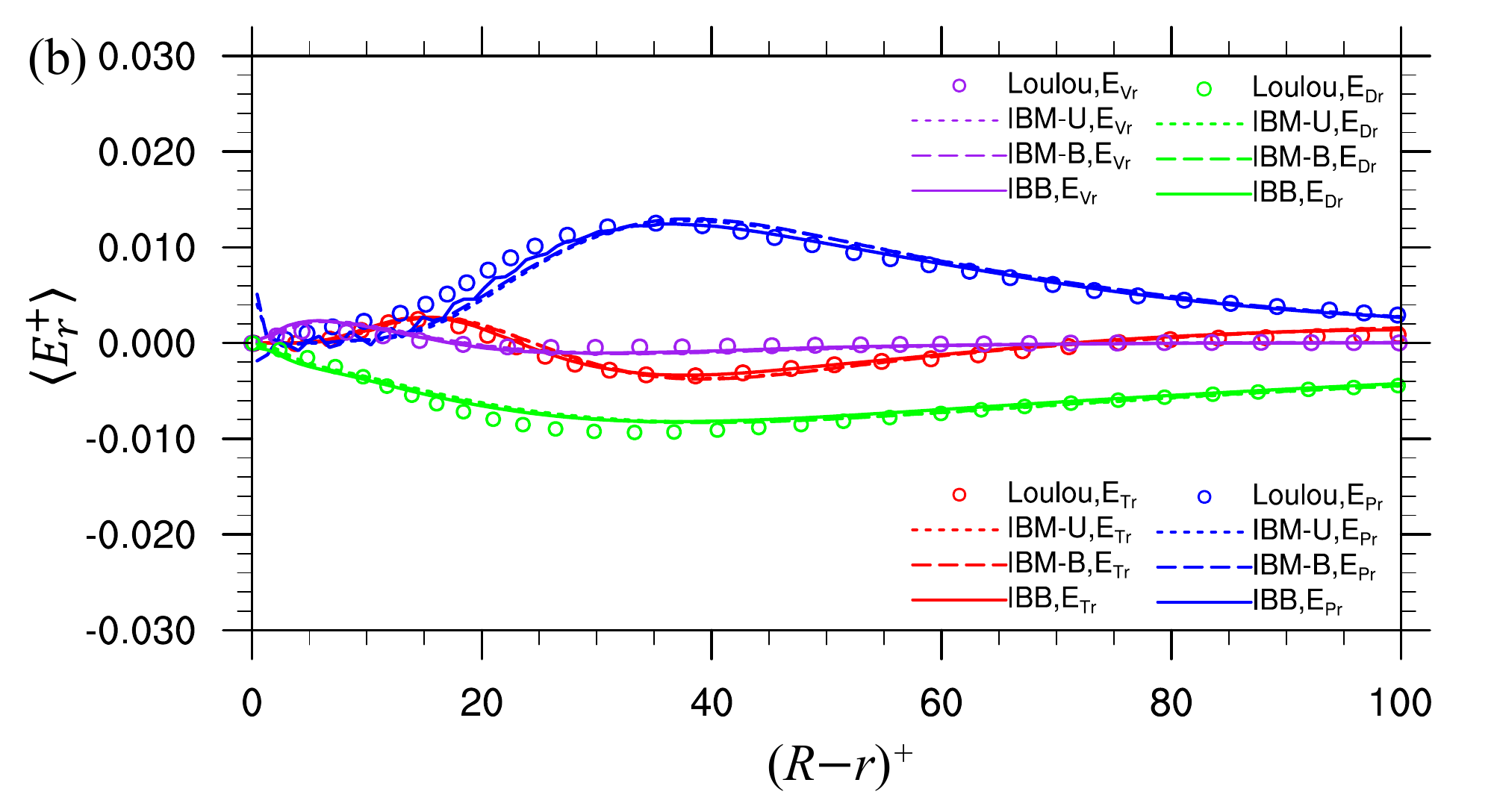}
\vspace{0.1in}
\includegraphics[width=110mm]{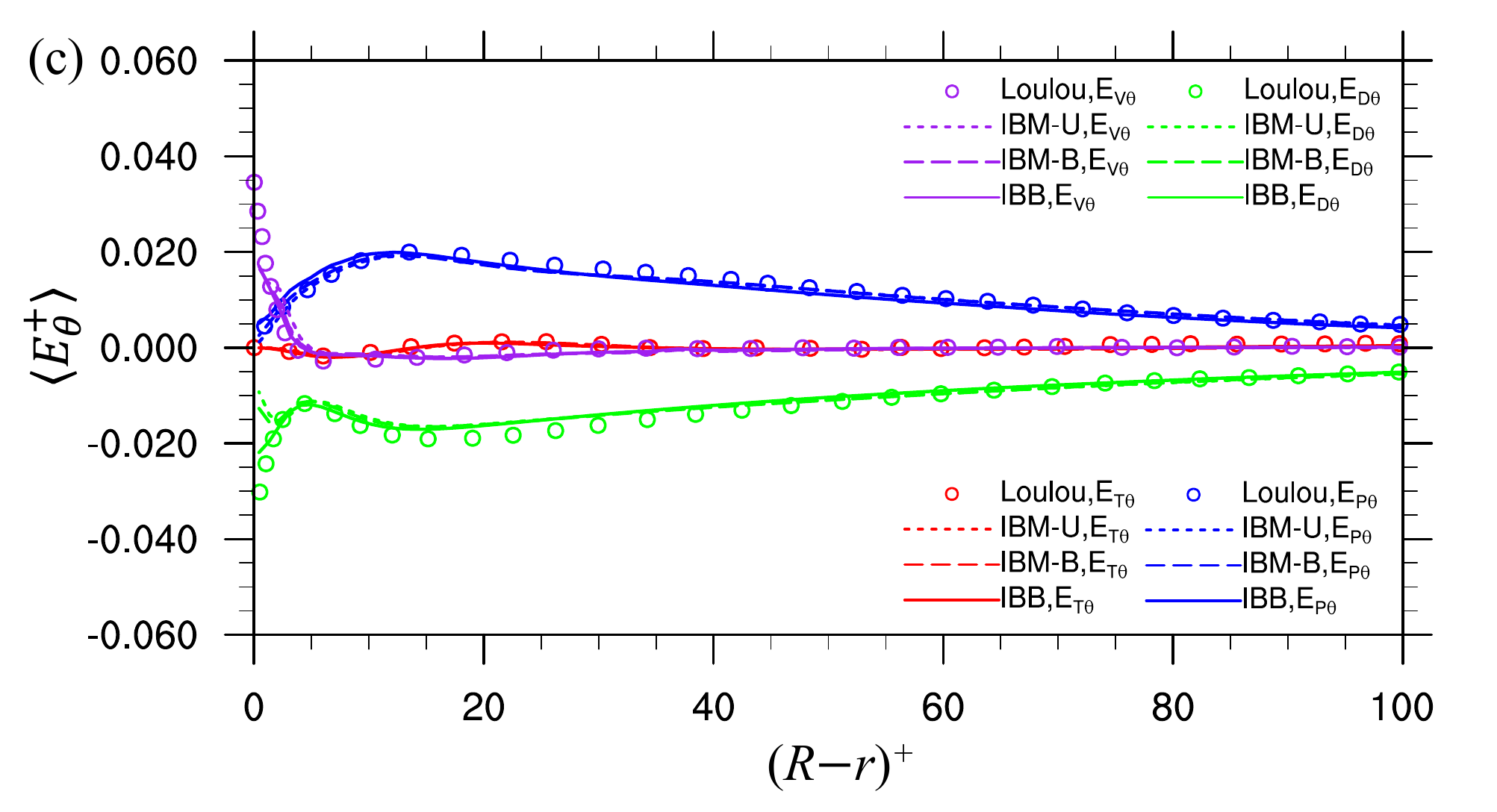}
\caption{Distribution of different terms in the budget analysis of: (a) streamwise TKE, (b) radial TKE, and (c) azimuthal TKE. Results from~\cite{Loulou97} are added for comparison.}
\label{fig:TKEbudget}
\end{figure}

$E_{Oi}$, $E_{Ti}$, $E_{Pi}$, $E_{Vi}$ and $E_{Di}$ represent the production, turbulent transport, pressure work, viscous diffusion, and viscous dissipation of TKE for the $i$-th velocity component, respectively. The flow fields at the stationary state  are used to compute each terms in the budget equations, and the results are compared with the benchmark results of the spectral simulation of Loulou {\it et al}.~\cite{Loulou97} in Fig.~\ref{fig:TKEbudget}. 

For most of the terms in the budget equations, the profiles of the IBB simulation and IBM simulations collapse with each other and match well with the benchmark results. However, for these terms involving the computation of velocity gradients and being non-zero at the wall, such as $E_{Vz}$, $E_{Dz}$, $E_{Pr}$, $E_{V\theta}$ and $E_{D\theta}$, the results from the two IBM simulations are clearly inaccurate for the same reason we mentioned earlier. The computation of local velocity gradients inside the diffused interface in IBM is a general problem of IBM, but not much attention was paid to this specific problem in the literature. The results from the IBB simulation, however, are much better because the interface is sharp and there is no boundary force added to change the governing equations solved in the simulation. 
 We also observe some unphysical fluctuations in the computation of term $E_{Pr}$ in the region of $(R-r)^{+}\le30$ in the IBB simulation. This probably results from the acoustic noises due to the intrinsic weak compressibility of LBM. Since the present simulations have a relatively high Reynolds number resulting in a small physical shear viscosity, the acoustic noises cannot be dissipated efficiently. A possible solution to this problem is to enlarge the bulk viscosity, by increasing the relaxation time of the energy mode in MRT LB models~\cite{luo2011numerics}, to enhance dissipation of the acoustic noises. The bulk viscosity does not appear in the incompressible Navier-Stokes equations, so its value is physically irrelevant to flow and can be optimized in the MRT LB model without impacting the physical results. Our code contains this feature but it was not activated to minimize the number of adjustable parameters. The current single relaxation parameter setting results in a bulk viscosity equals to the shear viscosity. The two IBM simulations, on the other hand, have no such problem, probably because the diffused interface in IBM introduces a numerical viscosity that helps suppressing the acoustic noises. The sharp interface treatment in IBB implies a smaller numerical viscosity. 

\section{Particle-laden decaying homogeneous isotropic turbulence}

\subsection{Problem description and simulation setup}

The second flow used for inter-comparison is a particle-laden decaying homogeneous isotropic turbulence, which is essentially the same case investigated recently by Br{\"a}ndle de Motta {\it et al.}~\cite{de2019assessment}. The physical parameters of the initial carrier turbulent flow are  identical to these given in Table 1 of~\cite{de2019assessment}. The initial turbulent flow field and particle positions are shared among all present simulations. However, for the particles, we examine the same particle size using two different grid resolutions, which is different from the set up used in~\cite{de2019assessment}. The parameters of the carrier flow and particles are recapitulated in Table~\ref{tab:parametersHIT}. The flow was simulated three times, one with IBB schemes and two with IBM algorithms. All three simulations used the same flow solver based on a D3Q19 MRT LB model. For this problem, a single relaxation time for all moments is no longer sufficient to maintain numerical stability, as the short-range particle-particle interactions further affect the numerical stability. We therefore used multi-relaxation times suggested by d'Humi\`eres {\it et al}.~\cite{dhumiers02} to help stabilize the simulations. In the IBB simulation (labeled ``PL-IBB''), the no-slip boundary condition on particle surfaces is realized by the quadratic IBB scheme of Bouzidi {\it et al.}~\cite{bouzidi2001momentum} by default. However, since particles could come very close with each other or even in physical contact, the default IBB scheme may not be applicable when there are not sufficient fluid grid points in the narrow gap between two nearby particles. In such specific occasions, the linear IBB scheme of Bouzidi {\it et al.}~\cite{bouzidi2001momentum} and the single-node second-order accurate bounce-back scheme of Zhao \& Yong~\cite{zhao2017single} takes over successively when the precedent one is no longer applicable. The hydrodynamic force and torque required to update the particle motion are evaluated by the Galilean invariant momentum exchange method (GIMEM)~\cite{wen2014galilean}. When a fresh fluid point is uncovered by a particle, the scheme proposed by Caiazzo~\cite{caiazzo2008analysis} was used to fill the distribution functions at this fresh fluid point. This procedure was also required in the IBB simulation for initialization since no fluid flow was made available in the solid region when the IBB scheme were used for the no-slip boundary treatment, but it was not required in the IBM simulations as a virtual flow field exist in the solid region when IBM was employed. 

The two IBM simulations both used the four-point delta function by Peskin~\cite{peskin2002immersed} to interpolate the velocity field from Eulerian mesh to Lagrangian mesh and the boundary force on the Lagrangian grid back to the Eulerian mesh. 
One of the IBM simulations uses Uhlmann's IBM~\cite{uhlmann2005immersed} (labeled ``PL-IBM-U'') and the other uses Breugem's IBM~\cite{breugem2012second} (labeled ``PL-IBM-B'') with a retraction distance of $r_{d} = 0.4$, same as in the turbulent pipe flow simulations in Sec.~2. For both IBM simulations, we calculate directly the fluid inertia inside the particles using the scheme proposed by Kempe et al.~\cite{kempe2012improved}. The boundary force in each simulation is also iterated twice after the initial prediction, to achieve more accurate no-slip condition on the particle surfaces.
We emphasize again that, although the original IBM algorithms of Uhlmann~\cite{uhlmann2005immersed} and Breugem~\cite{breugem2012second} are different in several aspects, the only difference we try to investigate here is how the retraction of Lagrangian grid points would affect the results of the simulated turbulent flow. 

For all three simulations, short-range hydrodynamic interactions or physical contact could happen frequently due to the large number of particles (or finite particle volume fraction). Typically, when two particles come very close, neither the flow inside the gap region between the two particles is fully resolved, nor the hydrodynamic interactions. A usual way to handle these scenarios is to introduce a lubrication model to handle the unresolved part of the hydrodynamic interactions so the correct particle motion is still predicted. There are multiple choices of the lubrication model. The repulsive barriers, {\it e.g.}, these proposed in Ref.~\cite{glowinski2001fictitious,feng2005proteus}, are frequently used in the particle-laden flow simulations, but they could not capture the realistic hydrodynamic interactions between two nearby particles. Another type of the lubrication model, {\it e.g.}, these proposed in Ref.~\cite{Ladd1997sedimentation,brandle2013numerical,costa2015collision}, are based on the theoretical lubrication force in the Stokes limit. Compared to the former, the latter category is more physical but involves adjustable parameters whose optimized values might depend on the no-slip boundary treatments. This is evident since the lubrication models are employed to supplement the resolved part of the hydrodynamic interactions, while the resolved part of the hydrodynamic interactions is largely determined by the no-slip boundary treatments. In order to avoid potential contamination due to the lubrication model, we simply drop out lubrication model in the present simulation. Instead, a soft-sphere collision model~\cite{brandle2013numerical} is employed to prevent particles from unphysical overlapping. This soft-sphere collision model only includes the normal contact force, the tangential solid-solid contact friction due to particle rotation were neglected.

Each particle was released into the initial HIT field with an initial translation velocity equal to the fluid velocity averaged over the volume occupied by the particle and a zero angular velocity. In the comparative study made by different groups,{\it i.e.}, Ref.~\cite{brandle2013numerical}, many of the discrepancies observed were due to particles initialization. In the present simulations, we have better control of the initialization procedure in all simulations to minimize the impact of initialization on the results. At the moment particles were inserted ($t = 0\,s$), the fluid volume occupied by the particles in the IBB simulation had no contribution to the hydrodynamic torque felt by the particles; but in the two IBM simulations, the flow inside particle volume did contribute. To reduce this difference in the initialization, we forced the initial velocity fields inside the particles to follow precisely the rigid body motion of the corresponding particles. This was an one-time correction at the moment when particles were inserted. The inner-particle flow fields in the two IBM simulations then evolve naturally. 

\begin{table*}
\caption{Parameters of the carrier flow and dispersed particles~\cite{de2019assessment}. The first table from the left to right: kinematic viscosity, Taylor length scale, Kolmogorov length scale, Kolmogorov time scale, initial root-mean-square flow velocity, initial eddy turnover time, Taylor Reynolds number. The second table from the left to right: grid mesh size, grid resolution, number of particles, particle/fluid density ratio, particle diameter per Kolmogorov length, particle diameter per Taylor length, particle diameter per grid spacing, particle volume fraction. }
\footnotesize
\begin{center}

\begin{tabular}
{c c c c c c c}
\toprule
$\nu [m^2/s]$&$\lambda[m]$&$\eta[m]$&$\tau_{K}[s]$&$u_{rms}^{0}[m/s]$&$T_{e}^{0}[s]$&$Re_{\lambda}$\\
\midrule 
$1.0\times10^{-3}$&$13.7\times10^{-2}$&$74.4 \times10^{-4}$&$55.2\times 10^{-3}$&$64.0\times10^{-2}$&0.8&87.6\\
\bottomrule
\end{tabular} 

\begin{tabular}
{c c c c c c c c}
\toprule
$N^3$&$k_{max}\eta$&$N_{p}$&$\rho_{p}/\rho_{f}$&$d_{p}/\eta$&$d_{p}/\lambda$&$d_{p}/\delta x$&$\phi_{p}$\\
\midrule 
512&1.90&4450&4.0&19.8&1.08&12.0&3\%\\
1024&3.81&4450&4.0&19.8&1.08&24.0&3\%\\
\bottomrule
\end{tabular} 

\end{center}
\label{tab:parametersHIT}
\end{table*} 

\subsection{Results and  discussions}

\begin{figure}
\centerline{
\includegraphics[width=180mm]{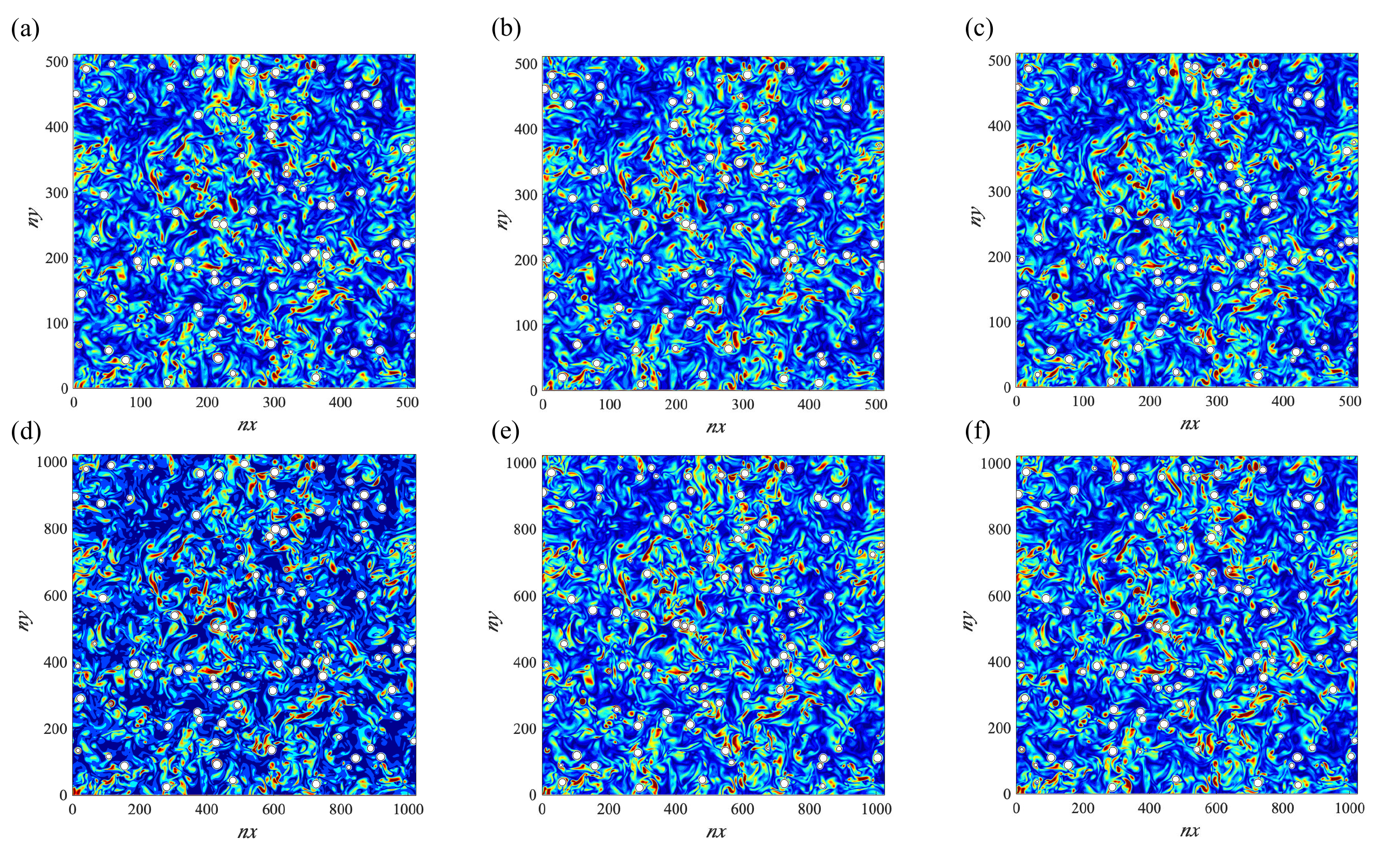}}
\caption{Contours of vorticity magnitude on an $x-y$ plane ($z=0$) at $t = 1.25\,T_{e}^{0}$. Contours in the first row are from three $512^3$ simulations, and these in the second row are from the three $1024^3$ simulations, from the left to right: PL-IBB, PL-IBM-U, and PL-IBM-B.}
\label{fig:visualizations}
\end{figure}

At $t = 1.25\,T_{e}^{0}=1\,s$, contours of vorticity magnitude for an $x-y$ plane ($z=0$) are shown in Fig.~\ref{fig:visualizations} for each simulation.  All the six simulations show very similar flow patterns and particle distributions at this relative small time, providing a cross validation for each simulation. As time is increased, the flow fields in different simulations will eventually diverge, due to the high nonlinearity of the turbulent particle-laden flow.   

\subsubsection{Flow statistics}

\begin{figure}
\centering
\includegraphics[width=90mm]{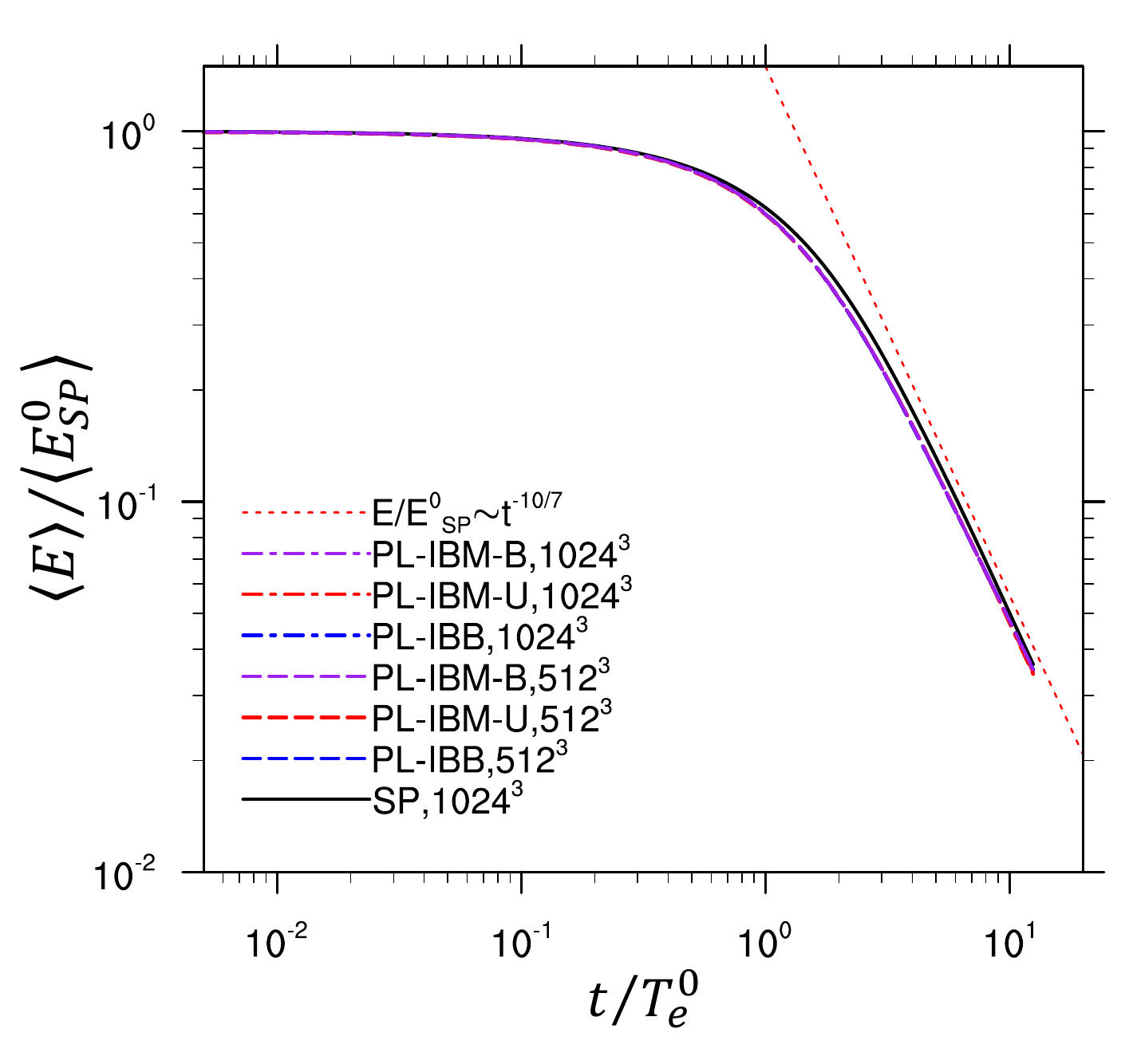}
\caption{Non-dimension averaged turbulent kinetic energy as a function of time.}
\label{fig:timedependentTKE}
\end{figure}

\begin{figure}
\centerline{
\includegraphics[width=90mm]{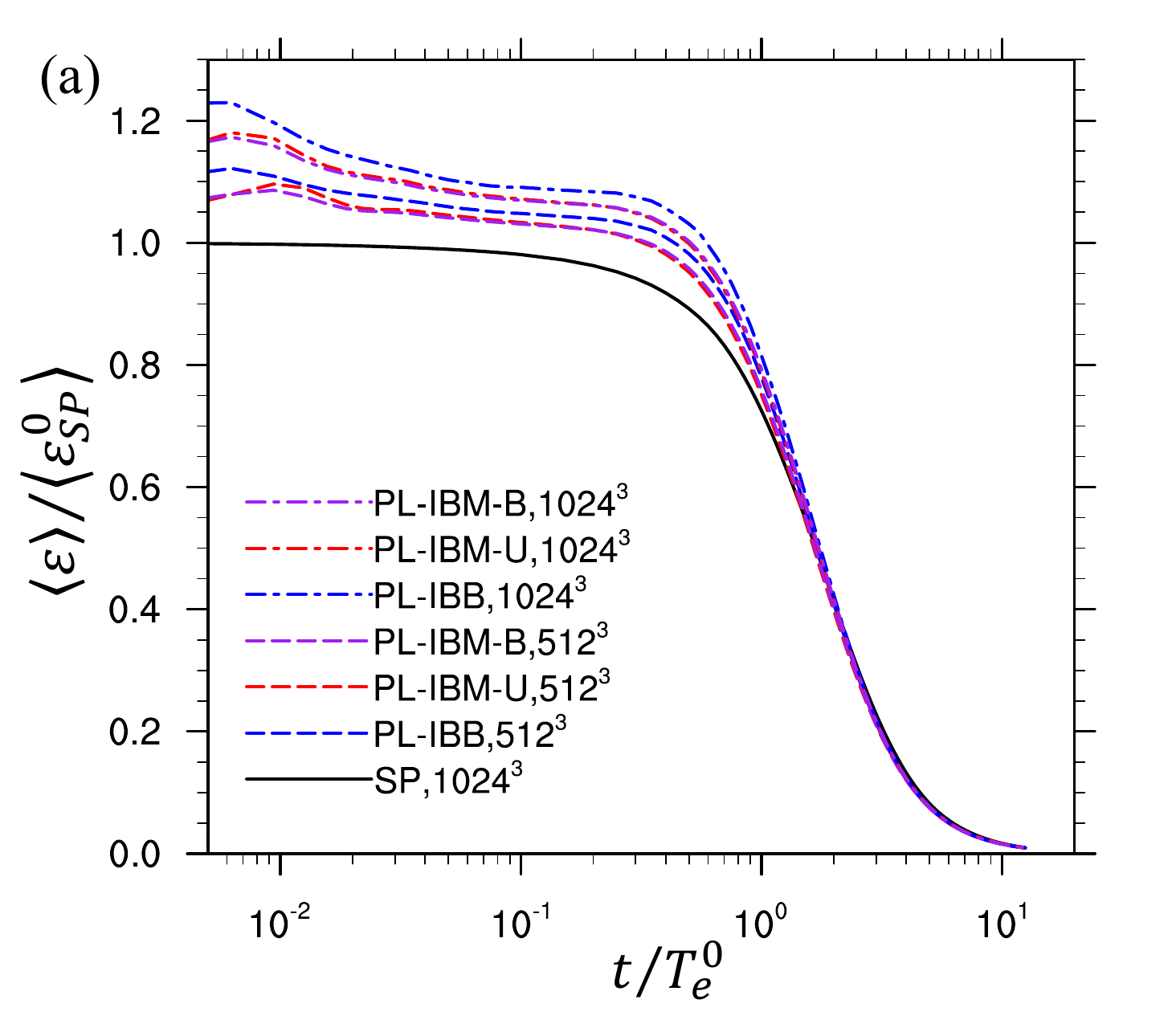}\includegraphics[width=90mm]{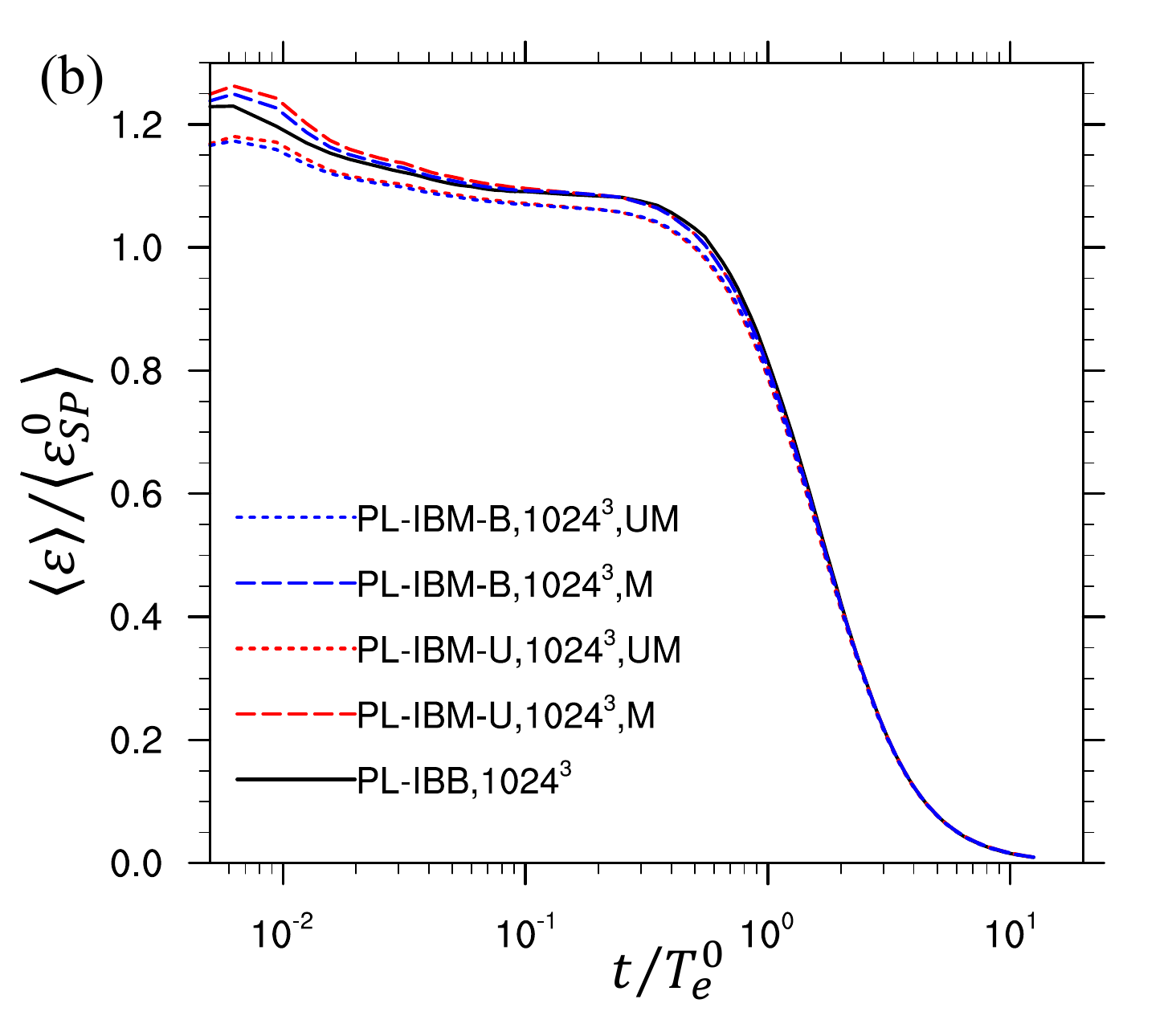}}
\caption{Non-dimension averaged dissipation rate as a function of time: (a) the comparison between IBB results and IBM results, the IBM results are obtained with masking the velocity field in the particle regions, (b) the comparison between masking and unmasking the velocity filed in the particle regions in the IBM simulations.}
\label{fig:timedependentdissip}
\end{figure}

We next examine the flow statistics. These flow statistics are computed in the Fourier space, except the local profiles. In order to enable Fourier transform, the velocity field must be continuous. In the IBB simulations, however, the fluid velocity is not available in the region occupied by the particles. To avoid discontinuity, we mask the region inside each particle with the velocity field generated from the rigid-body particle motion before the Fourier transform is conducted. In the IBM simulations, fluid covers the whole computational domain, so the velocity field inside the particles are already available. However, for fair comparisons, and for the sake of quantifying the contribution of the virtual flow inside particle regions, we conduct the statistics computation in the IBM simulations in both ways, {\it i.e.}, with and without masking the particle regions by the velocity field obtained from rigid-body particle motion. Note that in both IBB and IBM simulations, such masking is only a post-processing treatment, and it does not affect the flow evolution.

The time-dependent TKE and dissipation rates from different simulations are shown in Fig.~\ref{fig:timedependentTKE} and Fig.~\ref{fig:timedependentdissip}, respectively. The results ``SP"  in the two figures come from a single-phase LB simulation at the $1024^3$ grid resolution.   Fig.~\ref{fig:timedependentTKE} shows that all six simulations generated an almost identical time evolution of TKE, indicating that a grid resolution of $k_{max}\eta = 1.9$ is sufficient to capture the global TKE evolution, for both types of no-slip boundary treatments. The TKE of particle-laden flow decays slightly faster than the single-phase case due to the presence of particles. This faster decaying rate is due to the increased dissipation rate brought by particles at the early time, as shown in Fig.~\ref{fig:timedependentdissip}(a). For each boundary treatment, the simulation with the $512^3$ resolution  has slightly lower dissipation rate than the $1024^3$ simulation. This indicates that the grid resolution of $k_{max}\eta = 1.9$ is insufficient to accurately predict the dissipation rates in a particle-laden turbulent flow simulation.

 A quite well-known criterion of the grid resolution requirement for spectral DNS of single-phase HIT is $k_{max}\eta \ge 1.0$. Wang {\it et al.}~\cite{wang2014study} compared the grid resolution requirements of LBM and pseudo-spectral method (PSM) for DNS of single-phase HIT and concluded that the grid resolution for LBM should be doubled,  {\it i.e.}, $k_{max}\eta \ge 2.0$, relative to the grid resolution requirement for PSM. This criterion should also apply to other second-order finite-difference based methods as well. For particle-laden HIT, Wang {\it et al.} tested two grid resolutions, $k_{max}\eta = 3.08$ and $k_{max}\eta = 6.13$, and the two simulations resulted in similar flow statistics (when averaged over the whole domain), with only an exception of the flatness of velocity derivative (see Table.~2 in Ref.~\cite{wang2014study} for details). In the present simulations, a grid size of $1024^3$ corresponds to a grid resolution of $k_{max}\eta = 3.81$, which should be sufficient for most flow statistics if they are globally averaged. Unfortunately, due to the limitation of computational resources, we are not able to further increase the grid resolution to confirm this point. Compared to the IBB simulations, the IBM simulations with the same resolution under-predict the dissipation rate by roughly $2\%\sim3\%$. As we shall see later, this under-prediction of dissipation rate originates from the diffused interface in the IBM simulations, where the dissipation rate cannot be calculated correctly. 
 
 In order to quantify the impact of the post-processing, in Fig.~\ref{fig:timedependentdissip}(b) we compare the dissipation rate results from masking (M) and unmasking (UM) the particle regions with rigid-body motion in the IBM simulations. When the particle regions are masked, the contribution of the dissipation rate from the particle regions vanishes and it causes a visible drop of dissipation rate compared to the unmasking cases. Since the velocity field outside the particle regions is not affected by the masking, differences between two corresponding computations in Fig.~\ref{fig:timedependentdissip}(b) indicate the contribution of the virtual flow inside the particle regions in the IBM simulation. Physically speaking, this virtual flow should be viewed as a numerical error since it does not exist in reality. However, from the numerical simulation point of view, the existence of virtual flow in the IBM simulations does provide some additional dissipation rate that offsets the under-predicted dissipation rate within the diffused interface.
 
\begin{figure}
\centerline{
\includegraphics[width=80mm]{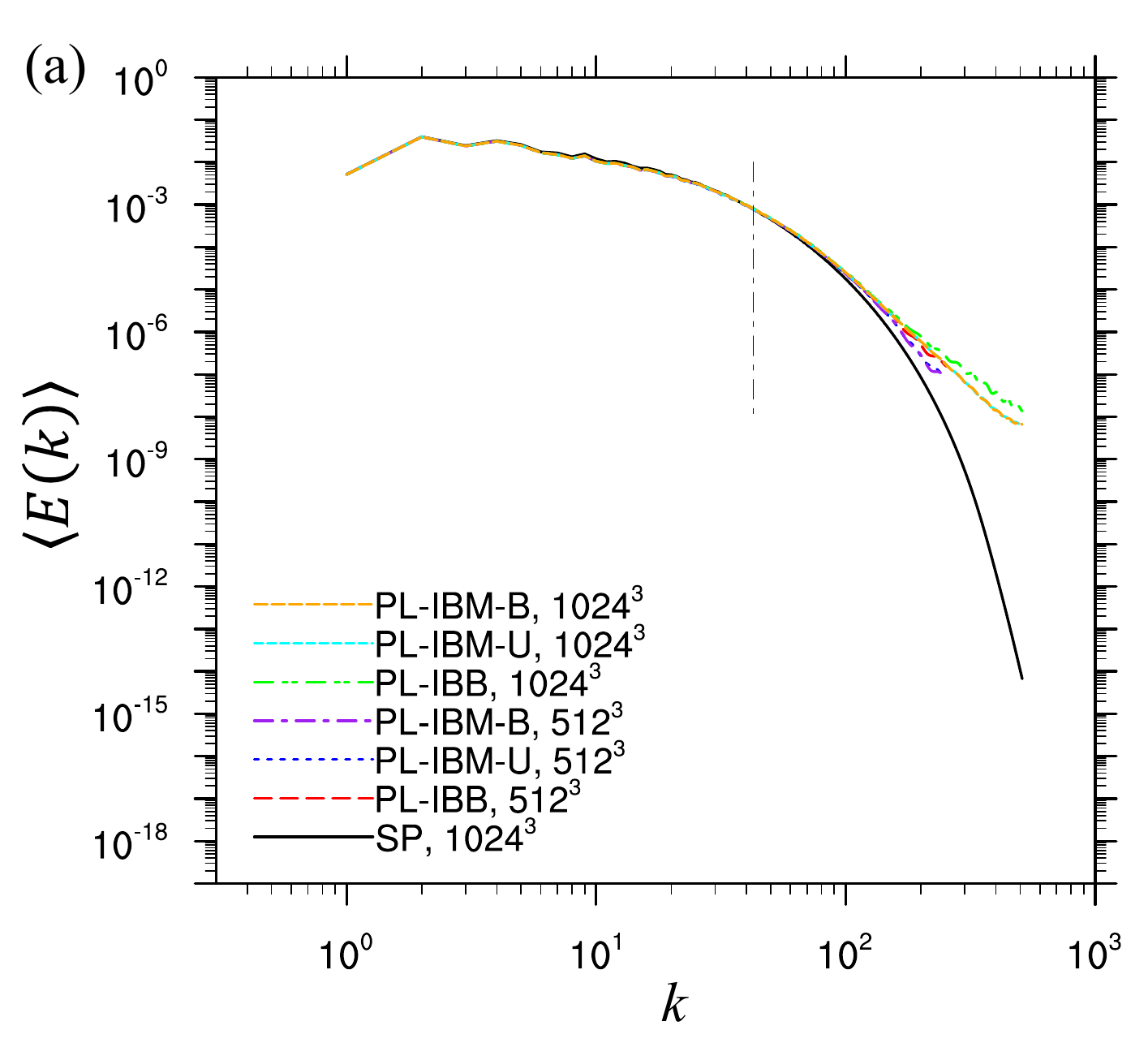}\includegraphics[width=80mm]{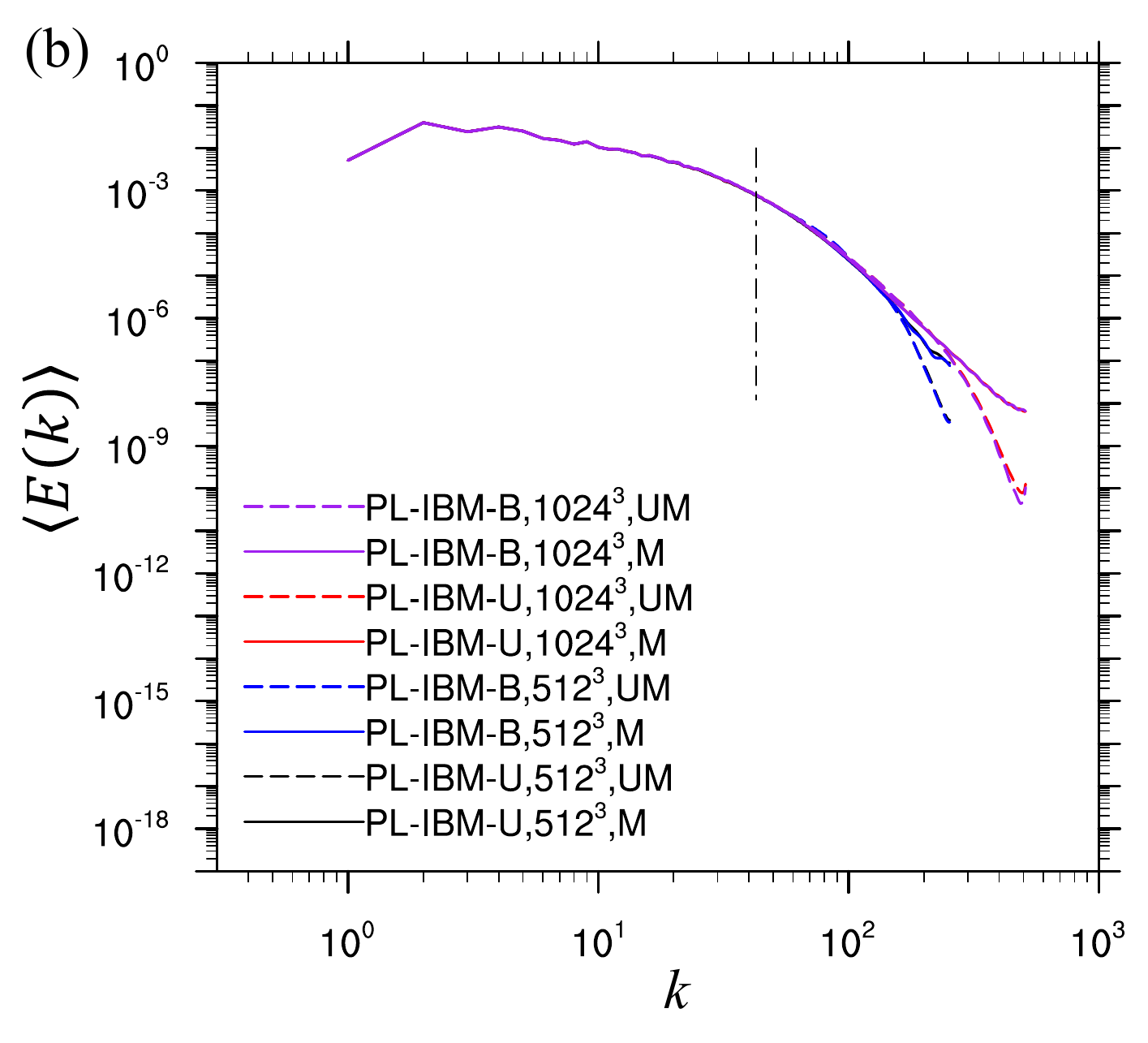}}
\vspace{0.1in}
\centerline{
\includegraphics[width=80mm]{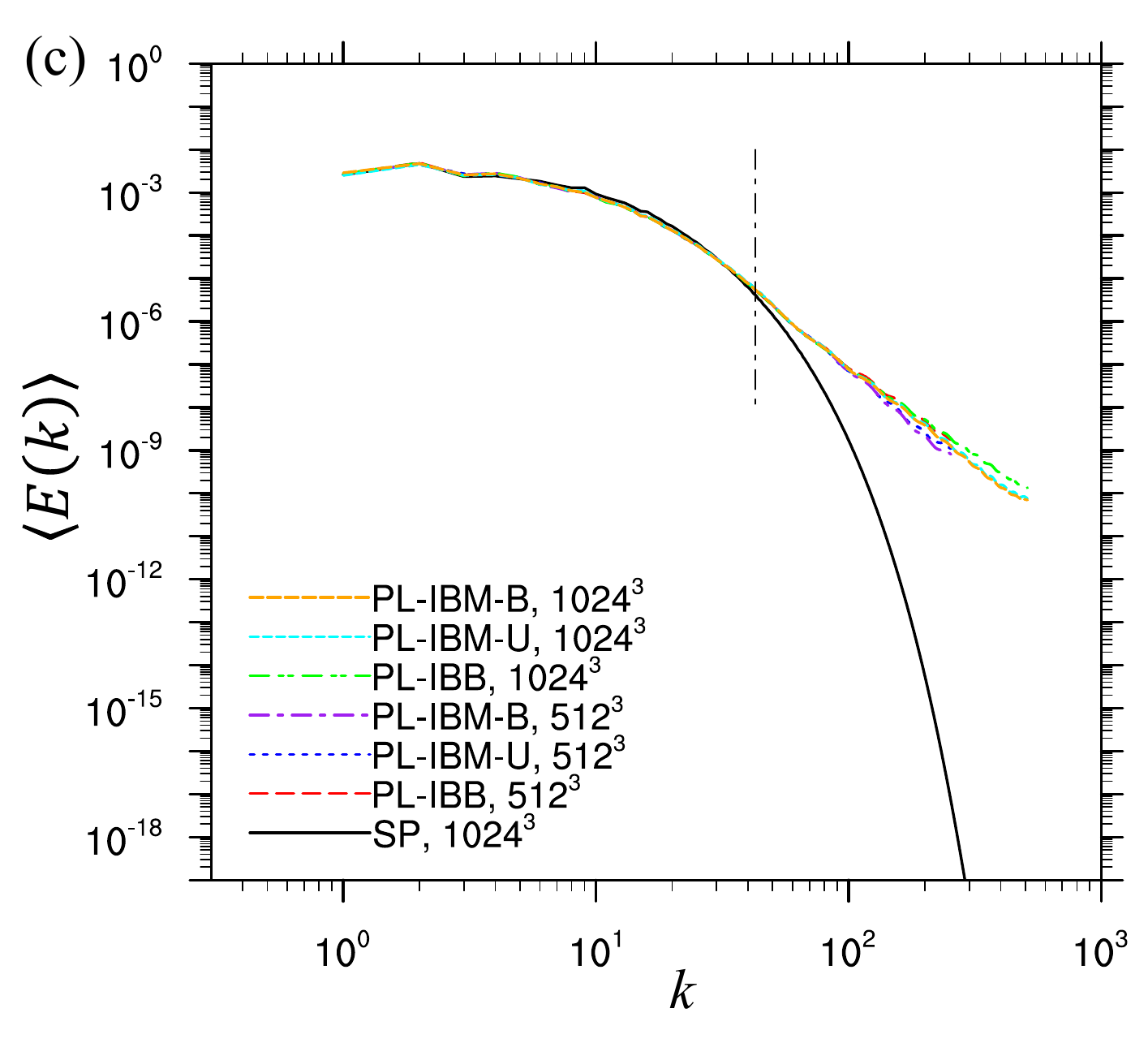}\includegraphics[width=80mm]{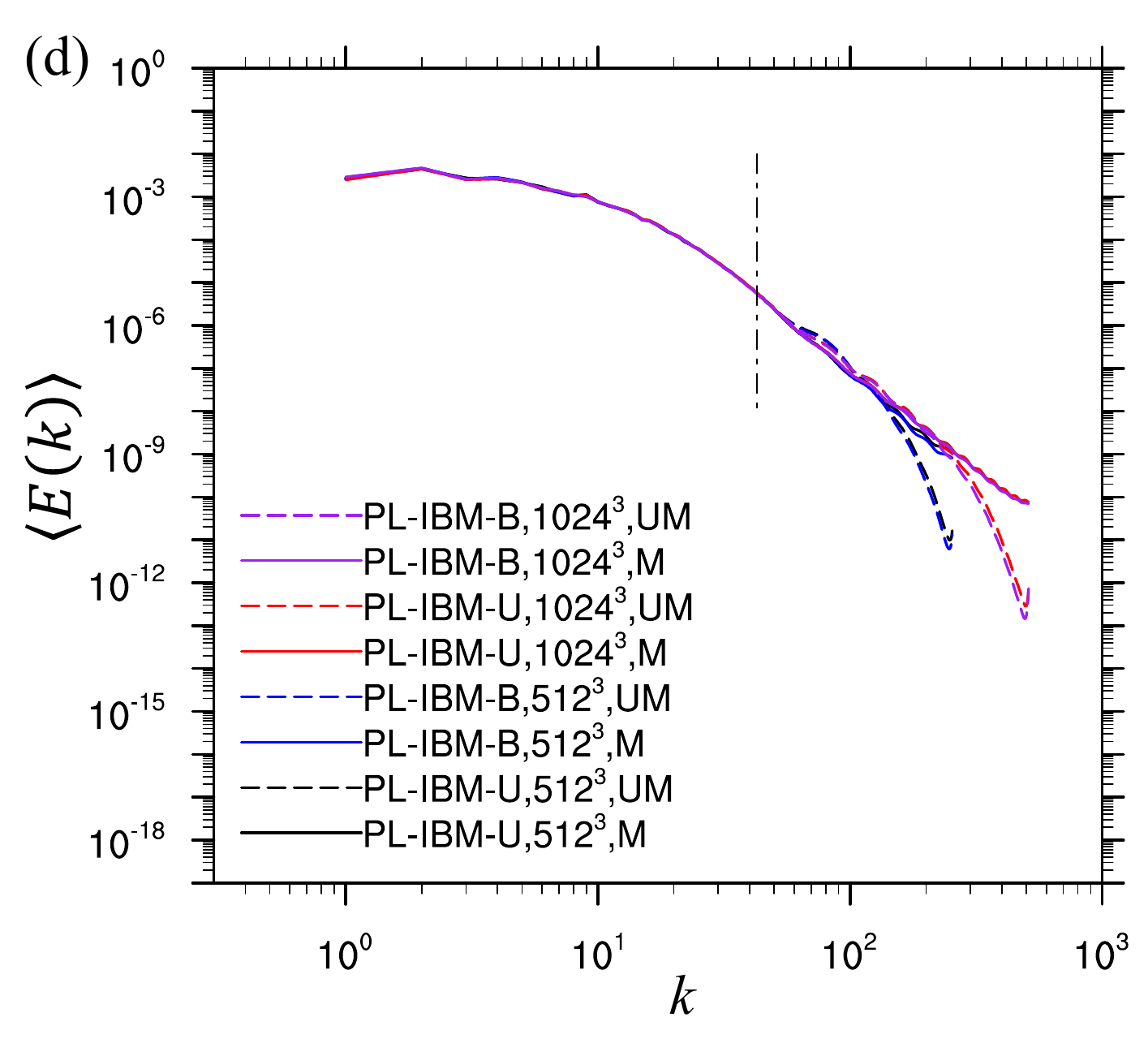}}

\caption{3D spectra of turbulent kinetic energy: (a) the comparison between IBB results and IBM results at time $t/T_e^{0} = 1.25$, the IBM results are obtained with masking the velocity field in the particle regions, (b) the comparison between masking and unmasking the velocity filed in the particle regions in the IBM simulations. The vertical line in each plot shows the wavenumber corresponding to the particle diameter, (c) same as (a) but at $t/T_e^{0} = 11.25$, (d) same as (b) but at $t/T_e^{0} = 11.25$. The vertical line in each plot shows the wavenumber corresponding to the particle diameter}
\label{fig:tkespectra}
\end{figure}

The three-dimensional spectra of TKE at two selected times $t = 1.25\,T_{e}^{0} = 1\,s$, and $t = 11.25\,T_{e}^{0} = 9\,s$ are shown in Fig.~\ref{fig:tkespectra}(a) and Fig.~\ref{fig:tkespectra}(c), respectively. When the velocity fields in the particle regions are masked, the TKE spectra of IBB and IBM simulations are identical for well-resolved large scales $(k\le 100)$ and only slightly different for smaller scales. Differences between two corresponding simulations with different grid resolutions are also small. The spectra from the two IBB simulations oscillate at very large wavenumbers. This could indicate the contamination of acoustic noises (typically on the scale of grid spacing). The IBM simulations, however, yield smoother spectra at these small scales, probably because the diffused interfaces help eliminate the high-frequency noises. While being more numerical dissipative, one may expect IBM to outperform IBB in terms of numerical stability, since the acoustic noises were previously found to be a potential source of numerical instability in LBM simulations~\cite{peng2016implementation}.

\begin{figure}
\centerline{
\includegraphics[width=90mm]{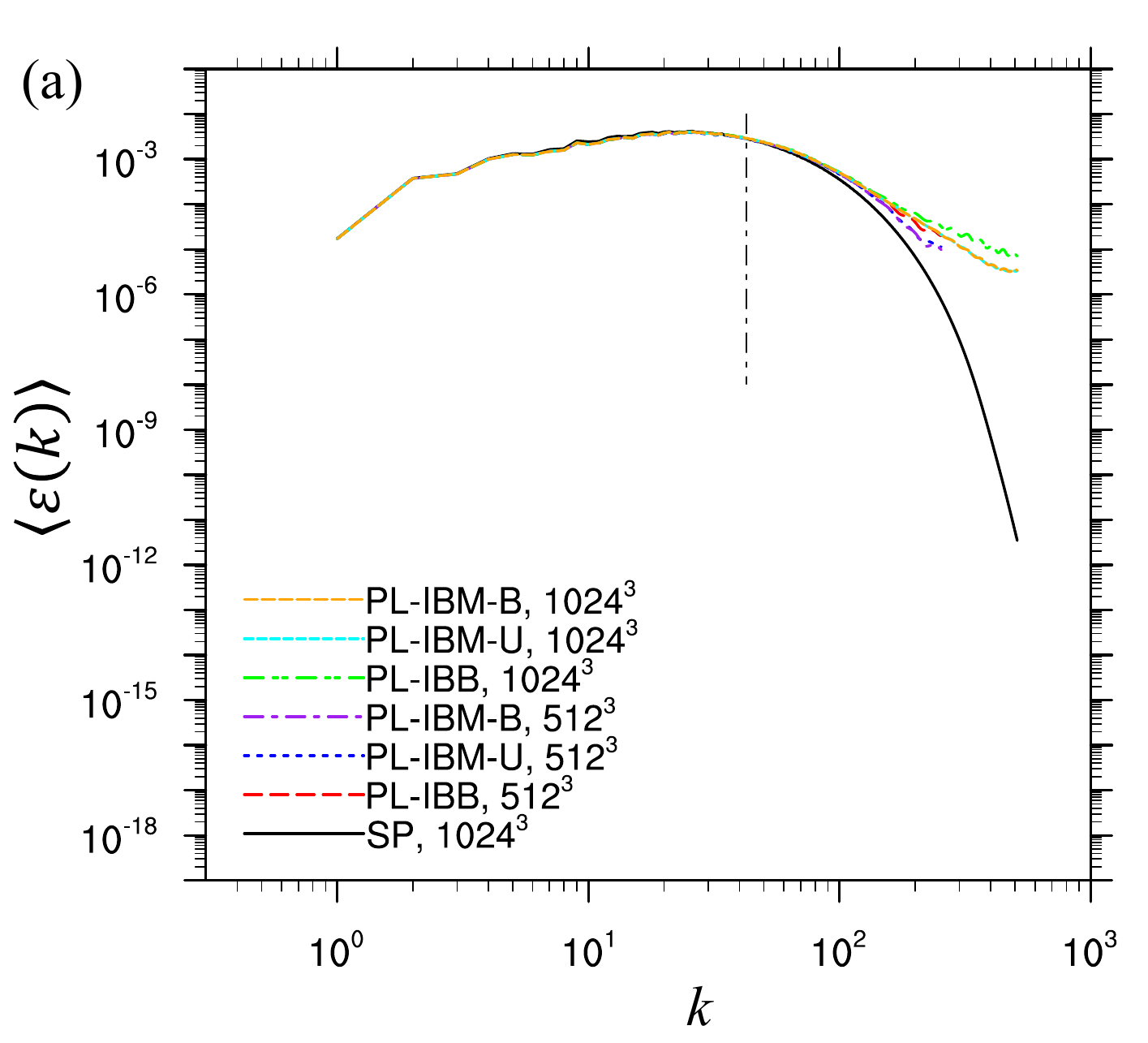}\includegraphics[width=90mm]{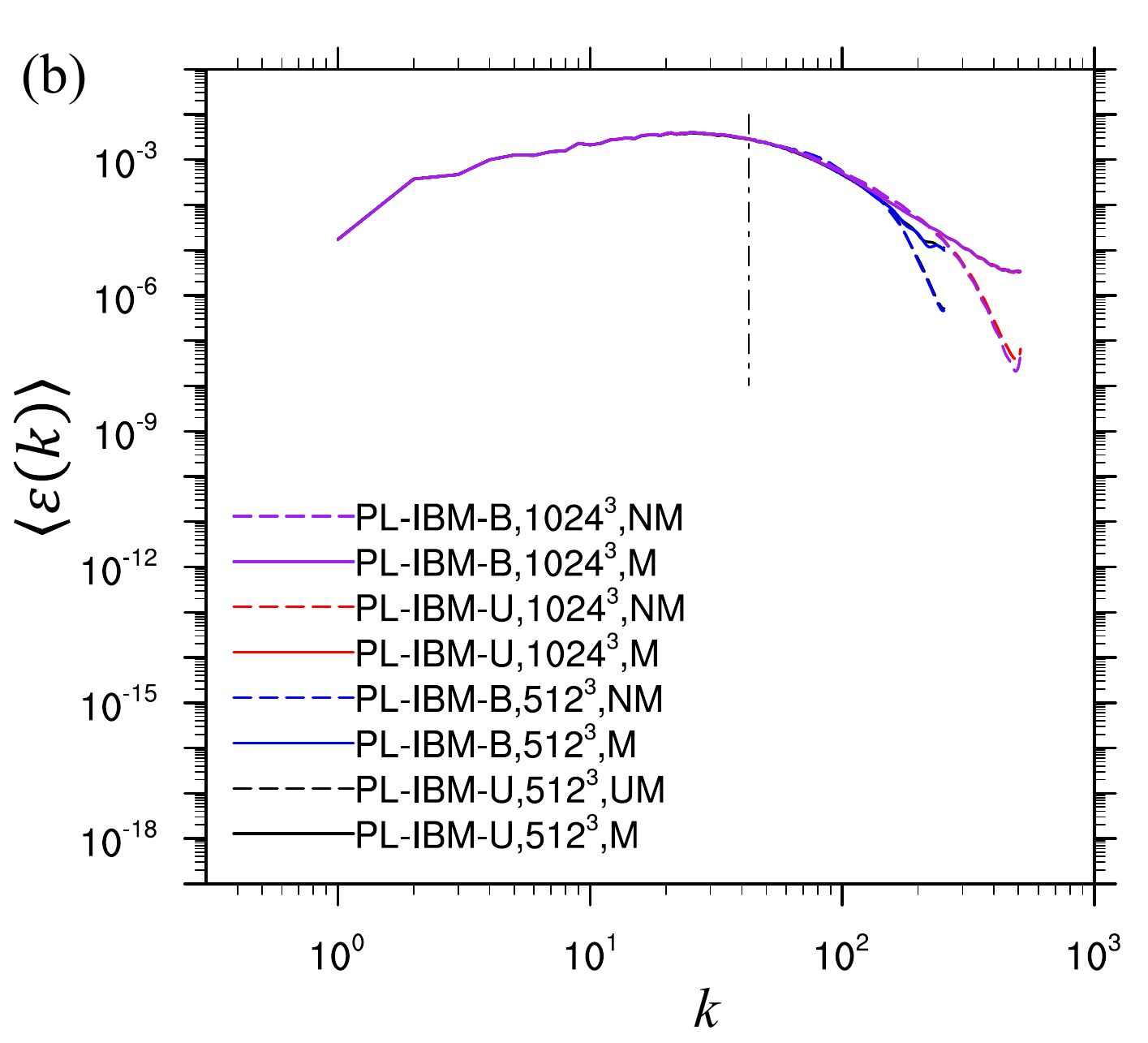}}
\vspace{0.1in}
\centerline{
\includegraphics[width=90mm]{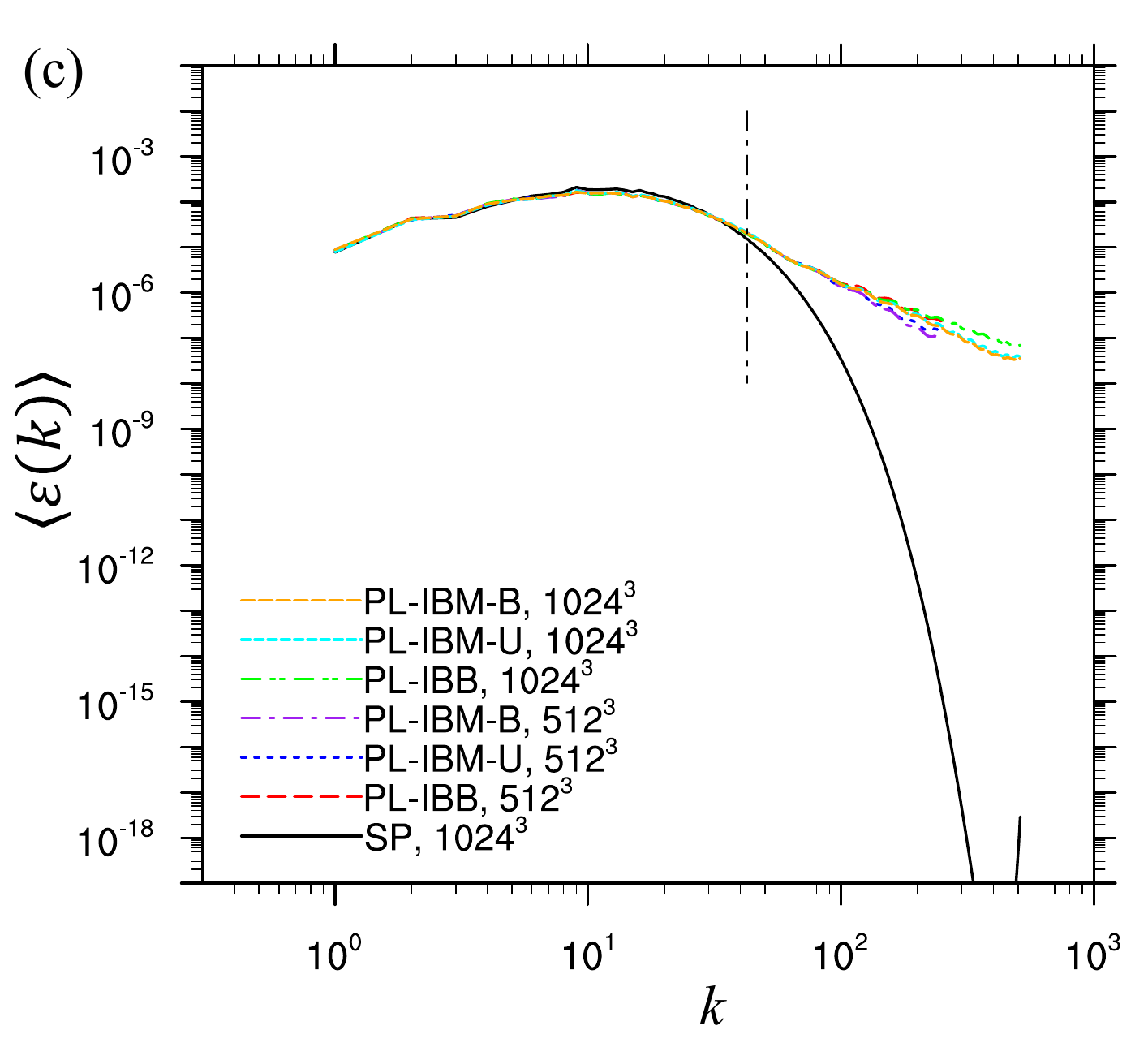}\includegraphics[width=90mm]{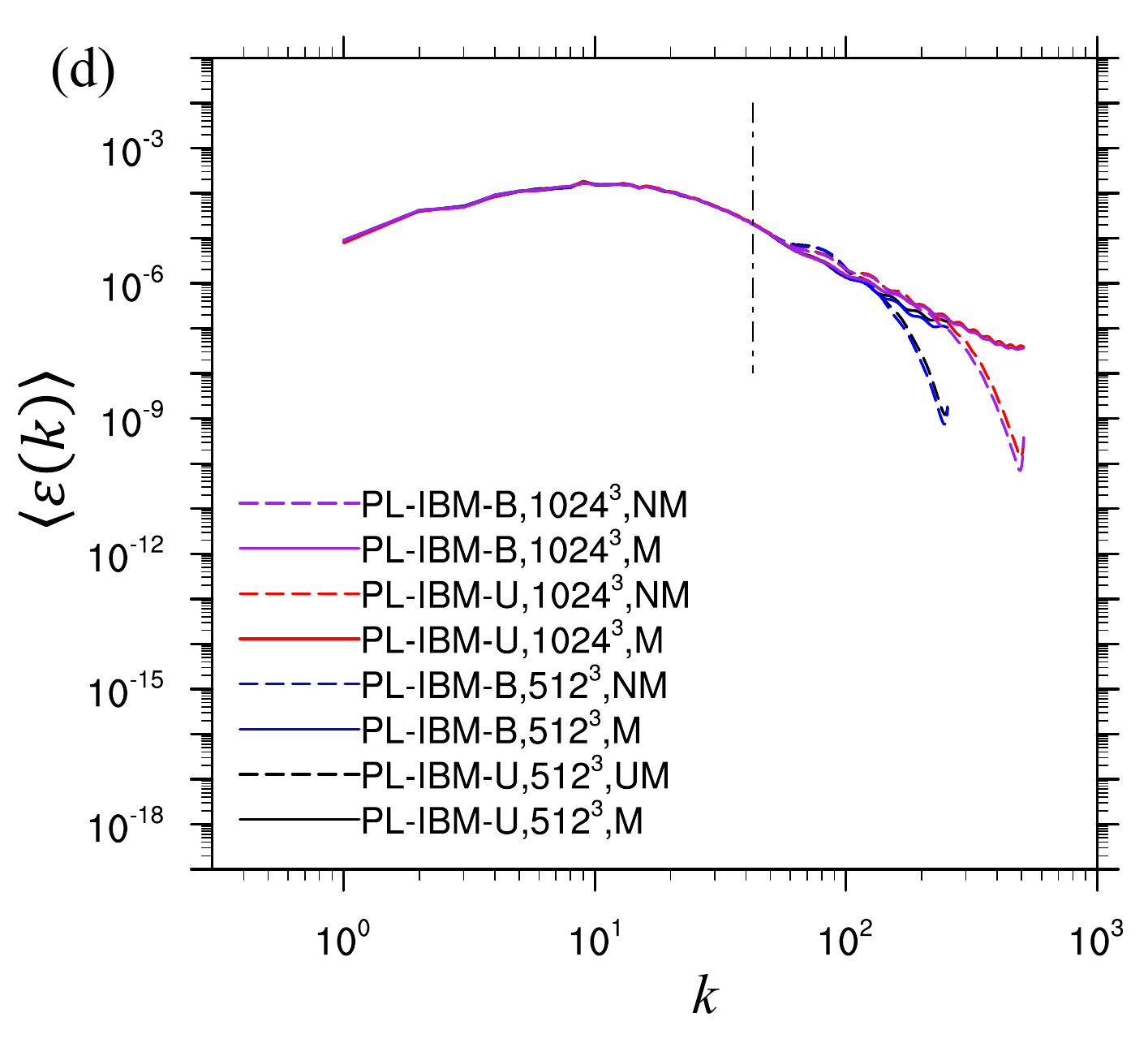}}
\caption{3D spectra of turbulent dissipation rate, same as Fig.~\ref{fig:tkespectra}.}
\label{fig:dissipationspectra}
\end{figure}

On the other hand, as shown in Fig.~\ref{fig:tkespectra}(b) and Fig.~\ref{fig:tkespectra}(d), whether  the velocity fields inside particle regions are masked  causes an obvious impact on the TKE spectra at large wavenumbers. Without masking the velocity fields, TKE is slightly overpredicted for wavenumber $50\lessapprox k\lessapprox 100$ but significantly underpredicted for the smallest resolved eddies compared to the results with masking. These derivations can be viewed as numerical errors contributed by the virtual flow inside the particle region, and they reduce at higher grid resolution. This is probably because when the grid resolution becomes higher, the forcing layer inside the particle occupies a smaller portion of the particle volume (0.704 for $512^3$ compared to 0.421 for $1024^3$). The virtual flow inside particles then becomes weaker and contributes to a smaller TKE derivation. The spectra of dissipation rate at the same two selected times are shown in Fig.~\ref{fig:dissipationspectra}. The comparison between the IBB simulations and the IBM simulations are similar to that of TKE spectra. Here we provide these results to complete the comparison.

\begin{figure}
\centerline{
\includegraphics[width=90mm]{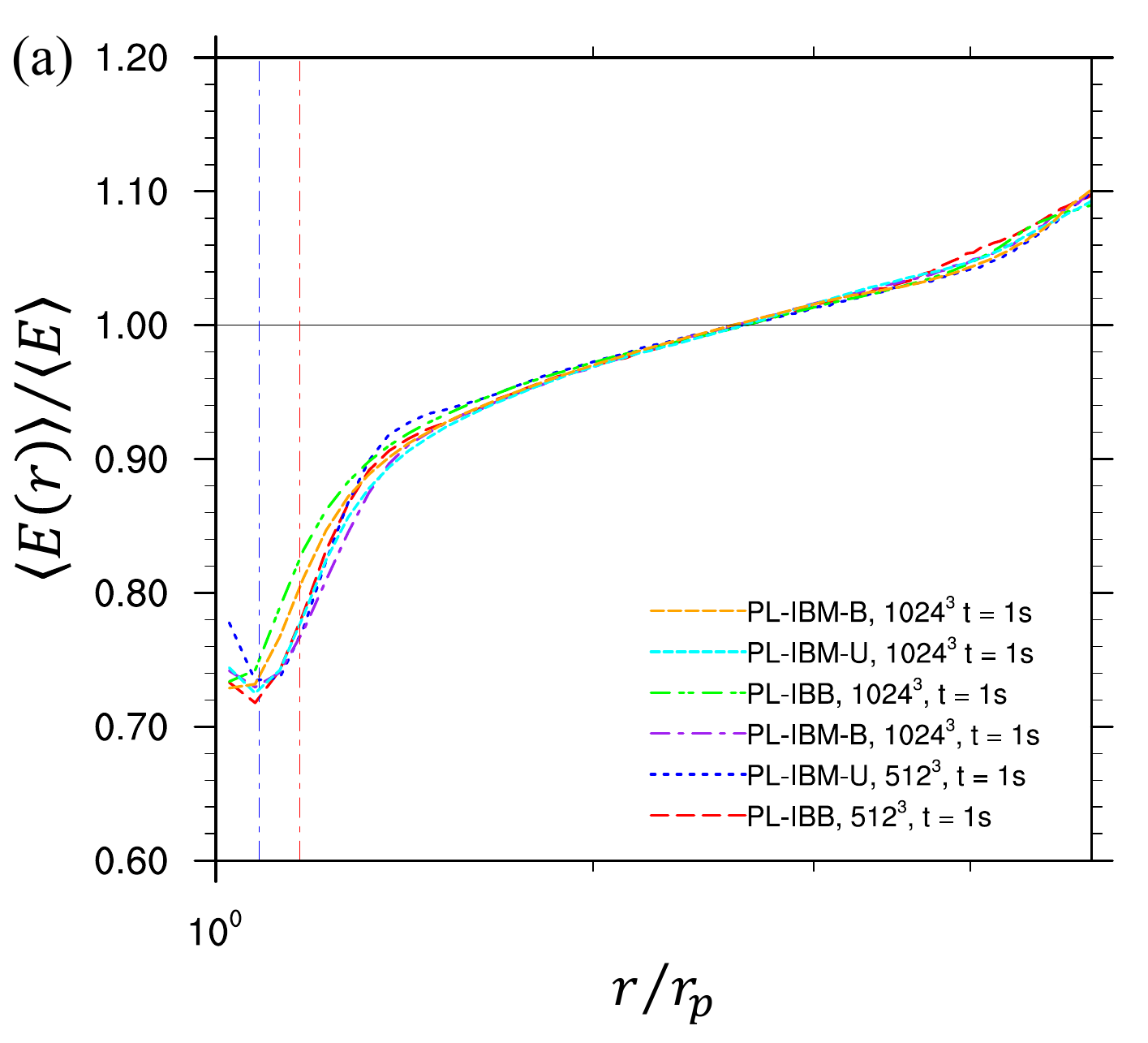}
\includegraphics[width=90mm]{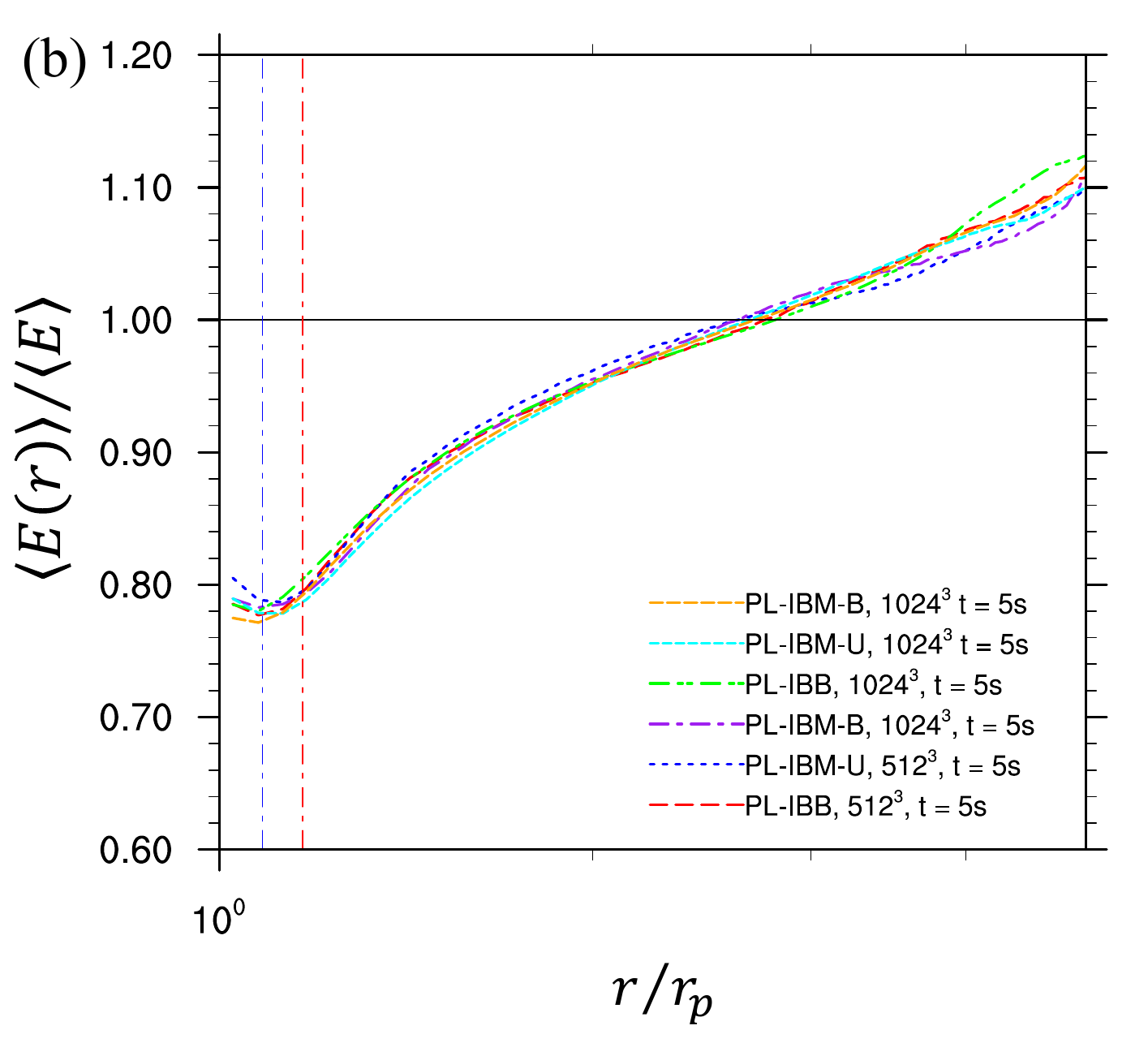}}
\caption{Profiles of TKE as a function of distance from the particle surface: (a) at $t/T_e^0 = 1.25$, (b) at $t/T_e^0 = 6.25$. The results are normalized by the averaged TKE over the whole fluid field at the same time. The two vertical lines represent the locations of $r_{p} + \delta x/r_{p}$ for each grid resolution.}
\label{fig:localtke}
\end{figure}

\begin{figure}
\centerline{
\includegraphics[width=90mm]{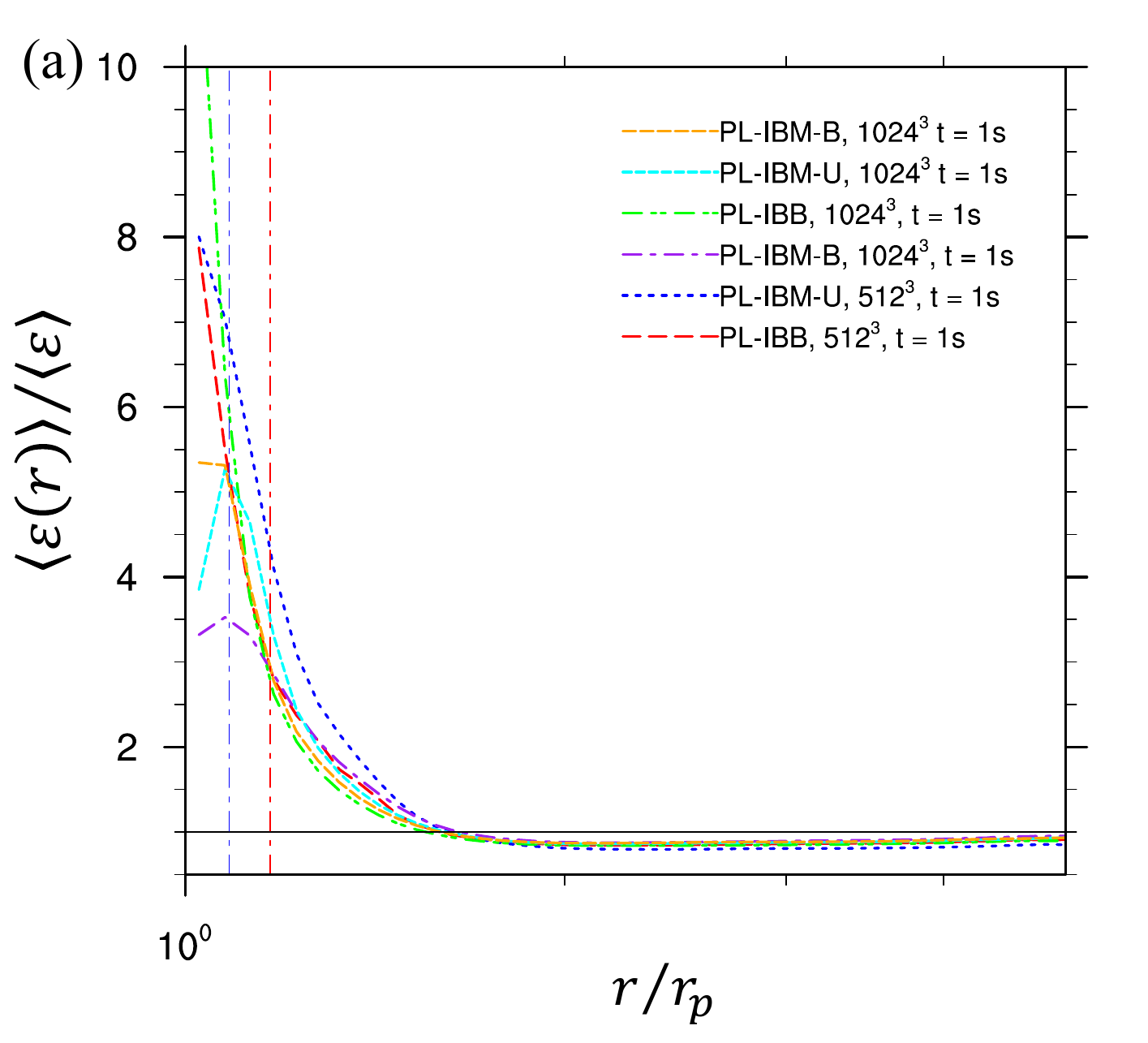}
\includegraphics[width=90mm]{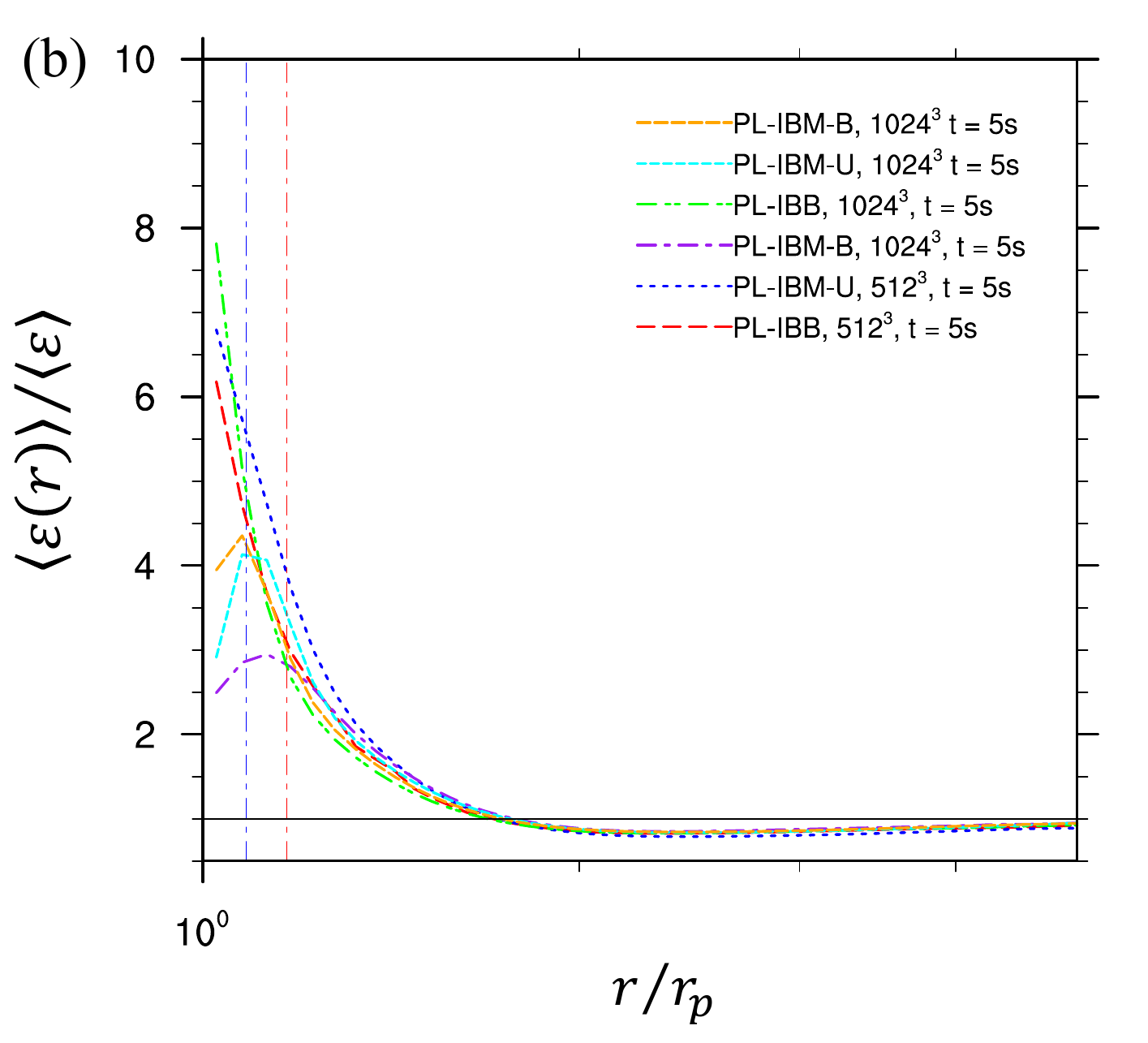}}
\caption{Profiles of viscous dissipation rate as a function of distance from the particle surface: (a) $t/T_e^0 = 1.25$, (b) $t/T_e^0 = 6.25$. The results are normalized by the averaged dissipation rate over the whole fluid field at the same time. The two vertical lines represent the locations of $r_{p} + \delta x/r_{p}$ for each grid resolution.}
\label{fig:localdissipation}
\end{figure}

We finish the flow statistics comparison between IBB and IBM simulations by comparing the locally averaged profiles of TKE and dissipation rate as functions of radial distance from the particle surface. The flow properties at each grid point are bin-averaged according to the radial distance from the nearest particle surface, so each grid point only contributes once for the computation. The grid points inside particles are excluded in this computation of local profiles. For simplicity, we assume the homogeneity and isotropy of the flow are not modified by the presence of particles, {\it i.e.}, the whole flow field still has zero mean, thus the TKE and dissipation rate of TKE are equal to the total kinetic energy and the dissipation rate of the total kinetic energy, respectively. This assumption has not been rigorously examined but it should be reasonable considering the large number of particles and their random distribution. 
The profiles of TKE and dissipation rate normalized by the field-averaged values as a function of distance from the particle surface ($r/r_{p} = 1$) are shown in Fig.~\ref{fig:localtke} and Fig.~\ref{fig:localdissipation} for two selected time frames $t = 1.25\,T_{e}^{0} = 1\,s$ and $t = 6.25\,T_{e}^{0} = 5\,s$. At $t = 1.25\,T_{e}^{0}$, the three simulations at $512^3$ all show a slightly decreasing TKE close to the particle surface followed by increasing TKE, as the distance from the particle surface is increased. With the higher grid resolution, the decreases of TKE close to the particle surface become less obvious. In the IBB and IBM-B simulations at $1024^3$ at $t=1.25\,T_e^0$, such decreases even disappear and TKE increases monotonically with the distance from the particle surface. Although we do not have direct evidence to support the reliability of the latter results, we do note that Burton \& Eaton reported a monotonic increasing of TKE in their simulation of a decaying HIT with a single fixed particle on a body-fitted grid mesh (see Fig.~16 in Ref.~\cite{burton2005fully}). It is also important to emphasize that in Burton \& Eaton's simulation, the particle was fixed, while in the present simulations, particles are allowed to move freely, so TKE at the particle surface is zero in the former but non-zero in the present simulations. 
At $t = 6.25\,T_{e}^{0}$, the profiles of TKE in the present simulations all slightly decrease then increase with increasing distance from the particle surface. If we use the simulations with higher resolution as the reference, the IBM-U simulation with $512^3$ appears to be the worst when compared to the IBB and IBM-B simulations at the same grid resolution. For globally averaged turbulence statistics (Fig.~\ref{fig:timedependentTKE} to Fig.~\ref{fig:dissipationspectra}), we did not observe significant improvements by retracting the Lagrangian grids, {\it i.e.}, ``IBM-B" compared to ``IBM-U". For local statistics, however, the retraction of Lagrangian grids does appear to improve accuracy. 

The effect of boundary treatments on the local TKE is relatively small. The impact on the results of local dissipation rate is more profound.
The local dissipation rates in the IBM simulations show drops approaching the particle surface, see Fig.~\ref{fig:localdissipation}. One can expect that strongest dissipation occurs at the particle surface due to the distortion and discontinuity brought by particles~\cite{eaton2009two}. The reduction of the dissipation rate near the surface observed with the IBM methods are, to our understanding, a numerical effect. Monotonic increase in the dissipation rate towards the particle surface was reported by in DNS of particle-laden HIT with body-fitted meshes with fixed particles, such as Fig.~17 in Ref.~\cite{burton2005fully} and Fig.~6(d) in Ref.~\cite{vreman2016particle} and for moving particles such as in Fig.~5(b) in Ref.~\cite{cisse_slipping_2013}. In the experiments of Ref.~\cite{tanaka_sub-kolmogorov_2010} (Fig.~19) the dissipation increases near the particle surface and there is no structured drop. The significantly underestimated dissipation rates near the particle surface in the IBM simulations indicate that IBM should be further improved or the dissipation rate should be properly redefined inside the diffused interface. Similar drops of dissipation rates near particle surface were reported in the literature. Lucci {\it et al.} showed drop of the dissipation in the back region of particles with a IBM approach (Fig.~15 to Fig.~17 in Ref.~\cite{lucci2010modulation}). Br\"{a}ndle de Motta {\it et al.} also observed this kind of drop with a VoF-Lag method (Fig.~8 in Ref.~\cite{brandle_de_motta_local_2016}). 
We hope that the current results, together with the results from the laminar flow tests in part I of this study, will alert the IBM community of the difficulties to analyze local dissipation rates near the immersed boundary.  
On the other hand, the two IBB simulations shows monotonically increasing dissipation rate profiles near the particle surface. This results seem more close to our physical expectations. However, the evident jump of dissipation rate in IBB with increasing grid resolution suggests that the grid resolution requirement to obtain accurate local profiles could be rather demanding, {\it i.e.,} a grid resolution of $k_{max}\eta = 3.81$ is not sufficient. The turbulent dissipation rate in the near vicinity of a freely moving particle remains an open question. This question has to be further quantitatively investigated experimentally, theoretically, and also numerically. Proper boundary treatments and post-processing treatments require further developments to ensure the accurate computation of the stress tensor near the particles.

\subsubsection{Particle statistics}

\begin{figure}
\centering
\includegraphics[width=80mm]{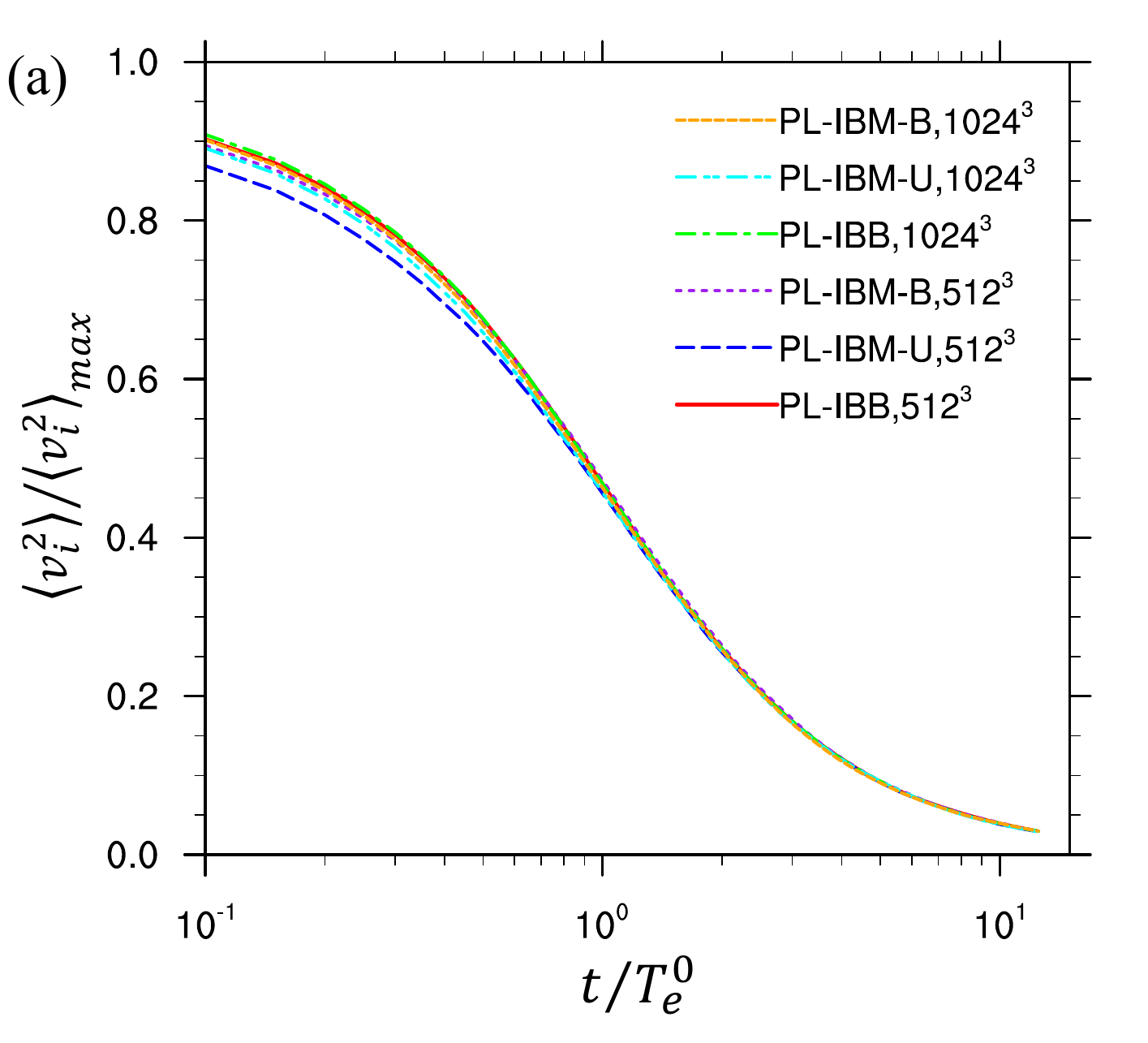}\includegraphics[width=80mm]{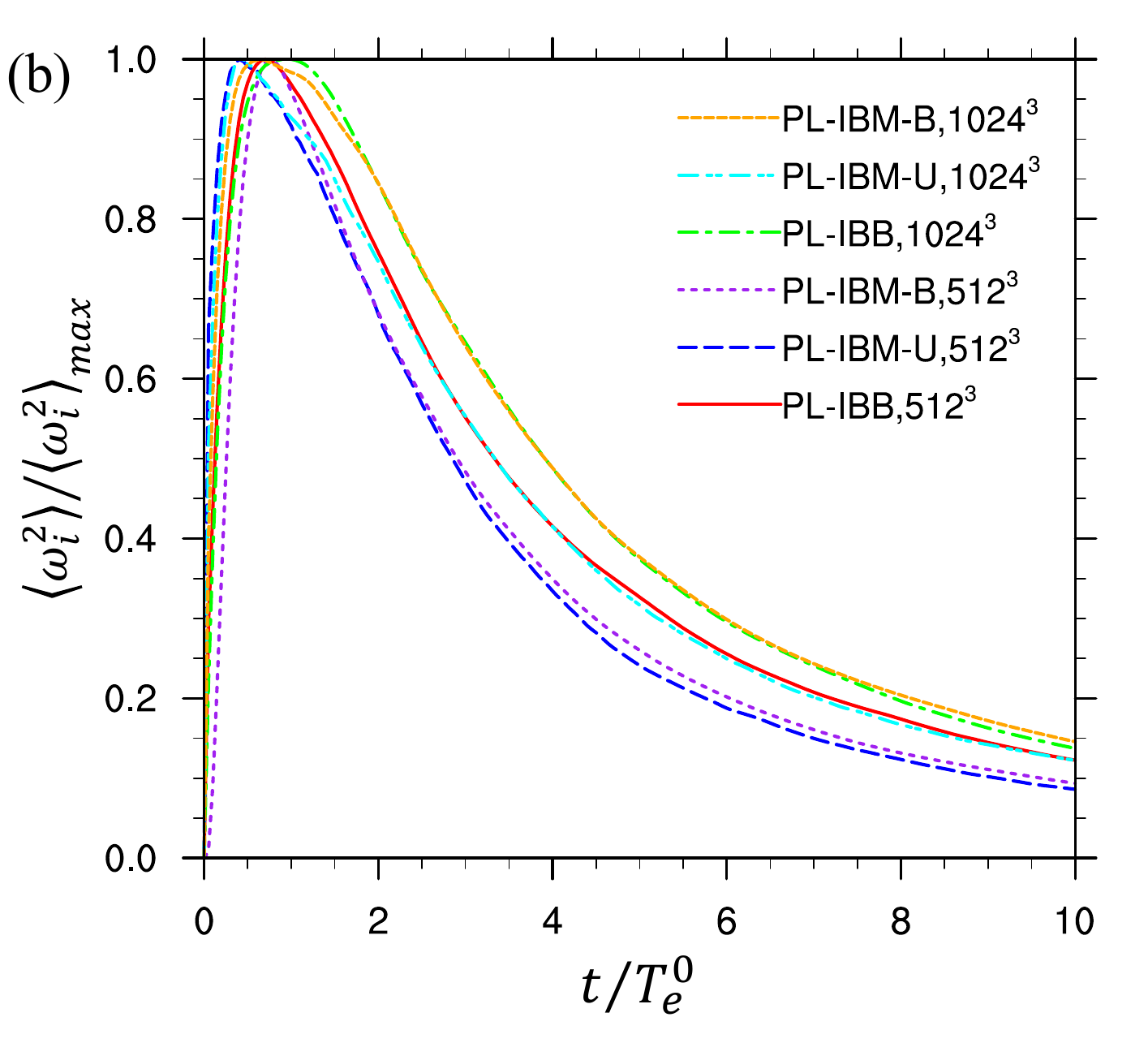}
\caption{Time evolution of (a) particle kinetic energy and (b) particle angular kinetic energy .}
\label{fig:particleTKE}
\end{figure}

Finally, we investigate the particle statistics. The time-dependent particle kinetic energy and angular kinetic energy are shown in Fig.~\ref{fig:particleTKE}. Compared to the fluid statistics, particle motion is more sensitive to the no-slip boundary treatment and the associated hydrodynamic force/torque evaluation. As shown in Fig.~\ref{fig:particleTKE}(a), the results of particle kinetic energy predicted from different simulations collapse well with each other, except a 4\% error in IBM-U with $512^3$ at the early times. This discrepancy shows the difficulty to implement consistent initial conditions among the different methods. The particle angular kinetic energy is zero initially since particles are released to in the fluid field with zero angular velocity. In all simulations, a maximum in particle angular velocity is reached around $t \approx T_{e}^{0}$, then decays afterwards. The results of particle angular kinetic energy show more discrepancies among different simulations. At $1024^3$, the results of particle angular velocity in the IBB simulation and the IBM-B simulation match well but different from the result of IBM-U. If we use the former two as reference, we observe that the results of particle angular kinetic energy from all three $512^3$ simulations deviate from the benchmark results quite obviously, but the IBB simulation has relatively better prediction of particle angular kinetic energy compared to the two IBM counterparts at later times ($t/T_{e}^{0} \ge 2$).

\section{Conclusions}
In this paper, we compared systematically the performances of two major categories of no-slip boundary treatment, the interpolated bounce-back schemes and the immersed boundary method in turbulent flow simulations within the lattice Boltzmann method. We hope the present investigation provides a more convincing assessment of these boundary treatments for turbulent flow simulations since direct inter-comparisons of these schemes in the context of turbulent flows are rare in the literature. 

In general, for the two turbulent flows investigated, {\it i.e.}, the fully developed turbulent pipe flow and the particle-laden decaying homogeneous isotropic turbulent flow, we found both categories of the no-slip boundary treatments can provide reliable results for most flow statistics. 
The major problem of IBM, as we already indicated in a series of laminar flow tests (see part I of this study in Ref.~\cite{peng2019}), is its inaccuracy in computing the local velocity gradients inside the diffused surface. This usually results in significantly underestimated local dissipation rate and viscous diffusion near the solid surface. The boundary treatment based on IBB schemes, on the other hand, is free from this problem. This is because when IBB is used, the fluid-solid interface remains sharp, and there is no virtual fluid field inside the volume occupied by the dispersed particles. The boundary treatment based on IBB is also found to be more accurate than IBM in capturing small-scale flow features near the grid scale. This is because the diffused boundary in IBM tends to eliminate these small scales. 

In the meantime, the diffused boundary in IBM can help suppress the acoustic noises due to the weak compressibility in LBM. Since these acoustic noises were found related to the numerical instability in LBM simulations, IBM is potentially more stable numerically than the IBB schemes.
The comparisons of results between the IBM-U simulations and IBM-B simulations show that the turbulent motion of the carrier flows are not sensitive to whether the Lagrangian grid points are retracted from the surface. However, we did observe a visible improvement of the mean flow velocity near the pipe wall with the retraction of Lagrangian grids in the turbulent pipe flow simulation. This indicates that the improved skin drag force prediction due to the retraction of the Lagrangian grids is mainly associated with the improved resulting mean flow. The retraction of Lagrangian points is found to improve the accuracy of the statistics of the dispersed particles as well. We therefore recommend IBM-B over IBM-U in LBM simulations.

While the flow solver used here is the lattice Boltzmann method, we believe the inaccuracy in computing the local velocity gradient is a general issue of IBM
within any flow solver. We would like to draw attention of this problem to the IBM community so that this issue can be addressed in the near future.
Finally, we would like to emphasize the importance of having a sufficient grid resolution in particle-laden turbulent flow simulations. For the particle-laden HIT investigated here, we found that a grid resolution of $k_{max}\eta \ge 3.8$ was necessary to ensure the accuracy of globally averaged dissipation rate. This grid resolution requirement is more critical when higher-order turbulence statistics and the local quantities near the particle surfaces studied. 

\vspace{10pt}

\begin{section}*{Acknowledgements}
This work has been supported by the National Natural Science Foundation of China (91852205 \& 91741101), and by the U.S. National Science Foundation (NSF) under 
grants CNS1513031 and CBET-1706130.
  Computing resources are 
provided by National Center for Atmospheric Research through CISL-P35751014, and CISL-UDEL0001. 
\end{section}



\bibliographystyle{elsarticle-num}

\bibliography{elsarticle-template.bib}

\end{document}